\documentclass[12pt]{article}
\pdfoutput=1
\usepackage{amsfonts,amssymb,amsmath,mathtools,cite}
\usepackage{float}
\usepackage{tikz}
\usepackage[dvipsnames]{xcolor}
\usepackage{hyperref}
\usepackage{orcidlink}
\definecolor{MyDarkBlue}{rgb}{0.15,0.15,0.45}
\hypersetup{
 colorlinks=true,
 linkcolor=MyDarkBlue,
 citecolor=MyDarkBlue,
 urlcolor=MyDarkBlue,
 pdftitle={Equivariant Localization with Electric Monodromy in M-theory},
 pdfauthor={Ji-Seong Chae},
 pdfsubject={hep-th}
}
\renewcommand{\baselinestretch}{1.2}
\numberwithin{equation}{section}

\newcommand{\ii}{\mathrm i}
\newcommand{\tr}{\mathrm{tr}}
\newcommand{\AdS}{\mathrm{AdS}}
\newcommand{\dd}{\mathrm{d}}
\newcommand{\vol}{\mathrm{vol}}
\newcommand{\ellp}{\ell_{\mathrm p}}

\begin{document}
\pagenumbering{Alph}
\begin{titlepage}
\vskip 1cm
\begingroup
\renewcommand{\thefootnote}{\ensuremath{\dagger}}
\begin{center}
{\Large\bf Equivariant Localization with Electric Monodromy in M-theory}
\vskip 1cm
{Ji-Seong Chae\,\orcidlink{0000-0002-1890-496X}\footnote{\href{mailto:jiseongchae17@gmail.com}{jiseongchae17@gmail.com}}}
\vskip 1cm
{\textit{Department of Physics, Hanyang University,}}\\
{\textit{Seoul 04763, South Korea}}
\end{center}
\endgroup
\vskip 0.5cm

\begin{abstract}
We extend equivariant localization for supersymmetric $\AdS_3$
backgrounds of M-theory to solutions with nonzero electric flux
periods. The associated local electric potential acquires additive
monodromy, contributing to the warped volume alongside R-symmetry
fixed points and outer boundaries. Using Killing-spinor identities,
we relate the electric periods to magnetic flux and, for a class of
compact backgrounds, the warped volume to the electric Maxwell
charge. We also derive a relation between the gravitational
anomaly and the electric flux. We apply these results to
$\AdS_3\times S^3\times S^3\times T^2$ and
$\AdS_3\times S^3\times K3\times S^1$, recovering their central
charges and deriving the R-current levels of the former directly
from eleven-dimensional supergravity. Finally, we study a noncompact
half-BPS $\AdS_3\times S^3\times S^3$ geometry fibered over a
periodic half-cylinder, where fixed-point, monodromy and
outer-boundary contributions coexist.
\end{abstract}

\end{titlepage}

\pagenumbering{arabic}
\pagestyle{plain}
\setcounter{page}{1}
\begingroup
\small
\renewcommand{\baselinestretch}{1}
\selectfont
\tableofcontents
\endgroup
\newpage

\section{Introduction}\label{sec:intro}
Supersymmetric $\AdS_3$ backgrounds of string and M-theory
provide holographic descriptions of two-dimensional
superconformal field theories. Their leading central
charges follow from the Brown--Henneaux relation and
the internal warped volume \cite{BrownHenneaux}, while
quantized fluxes, R-symmetry and anomalies constrain the
dual theories \cite{MSW,Gauntlett2001,Gauntlett2006,
GauntlettAdS3,BeniniBobevExact,BeniniBobev,KrausLarsen}.

Equivariant localization replaces suitable bulk integrals
by fixed-point data of a symmetry action
\cite{BerlineVergne,AtiyahBott}. Its supergravity
formulation extracts the relevant closed forms from
supersymmetry and has been applied to holographic actions,
wrapped branes, volumes and black-hole observables
\cite{BenettiGenolini2019,BGS2023,BGSAdSCFT,
MartelliZaffaroni,BGSWrapped,ColomboVolume,SuhWrapped,
BGSAttraction,BGSFree,Cassani2024}. Related developments
cover toric and gauged backgrounds, punctures,
ten-dimensional theories and Chern--Simons couplings
\cite{BGSToric,BGSD4,CouzensRomans,BGSD5,
CouzensPunctures,CouzensTen,ColomboOdd,BGOdd2026,MTheoryPath}.

{%
For supersymmetric $\AdS_3\times M_8$ backgrounds,
the internal Killing-spinor equations determine the
R-symmetry vector and its bilinears \cite{Ashmore}.
Gregory, Notman and Sparks (GNS) used these bilinears to
construct equivariantly closed forms whose top components
are the magnetic flux, the M2-brane calibration form and
the warped volume \cite{GNS}. The lower components of these
forms contain a scalar potential for the internal electric
one-form. GNS use this potential globally, which requires
the electric one-form to be exact \cite{GNS}.

The electric one-form is closed by the Bianchi identity,
but it can have nonzero periods. Local scalar potentials
then differ by constants around electric cycles, and we
call this additive shift \emph{electric monodromy}. Its
effect on localization is sharpest when the R-symmetry acts
freely on a compact internal space. The fixed-point set is
then empty, so the fixed-point theorem applied to a global
equivariant completion of the warped volume would give zero.
The warped volume, and with it the leading central charge,
is nevertheless positive. This paper asks how the central
charge localizes in this situation, and what the electric
period means physically.

We show that the warped volume obeys
\begin{equation*}
 \int_{M_8}\mathrm e^{9\lambda}\vol_8
 =I_{\mathrm{fixed}}+I_{\mathrm{monodromy}}+I_{\mathrm{outer}}
\end{equation*}
on compact internal spaces and on regulated noncompact
domains. To obtain this formula, we construct
from the local GNS forms a seven-form $Q_7(y)$ satisfying
$\dd Q_7(y)=\mathrm e^{9\lambda}\vol_8$ away from the
R-symmetry fixed set. We remove small neighbourhoods of
that set and cut the space into regions on which the
electric scalar $y$ has a single branch. Stokes' theorem
then converts the warped volume into integrals over the
resulting boundaries. The shrinking neighbourhoods give
the GNS fixed-point residues, the two faces of each cut
give the jump of $Q_7(y)$, and any outer boundary supplies
the third contribution. If all electric periods vanish,
the cut jumps disappear and we recover the GNS formula.

The electric period is itself a physical quantity. The
Killing-spinor equations relate it to the magnetic flux
through the two-form bilinear that completes the magnetic
four-form. For compact backgrounds with a single electric
period and an $SU(3)$-structure spinor, the cut term is the
product of the period and the electric Maxwell charge, and
the Page charge differs from it by a magnetic term
\cite{Marolf2000,GautasonVanMuiden2025,Adhikari2026}. The
same electric class decides whether the magnetic flux has a
single global ordinary equivariant completion, and it
determines the $C_3\wedge X_8$ contribution to the
gravitational anomaly
\cite{DuffLiuMinasian,WittenFivebrane,FreedHarvey,
ChowPang,GregoryAnomaly}. A nonzero anomaly contribution
therefore detects the obstruction to that completion. The
localization formula holds in every case, since it uses
only the local GNS forms and their jumps across the cuts.

Three solutions illustrate these statements. On the
intersecting-brane background
$\AdS_3\times S^3\times S^3\times T^2$
\cite{Boonstra1998,Gukov2004,HansenKraus,HST2025} the
R-symmetry acts freely and the cut supplies the entire
warped volume. The electric period also follows from the
quantized magnetic flux, and the bulk R-current levels
reproduce the same quadratic dependence on the trial
R-symmetry. On $\AdS_3\times S^3\times K3\times S^1$ the
$C_3\wedge X_8$ contribution is nonzero and detects the
magnetic obstruction.
This product background is a constant-warp-factor
specialization of the Calabi--Yau two-fold family
\cite{LozanoMacphersonNunezRamirez2019,LozanoNunezRamirezSpeziali2020}.
Our periodic realization of the known half-BPS
three-sphere fibration
\cite{DHEGKLocal,DHEGKGlobal,BachasDHEK2014,
GentleGutperleMarasinou,ContiMacpherson2024}
has fixed points, an electric cut
and an outer boundary at once, and its regulated volume is
compared with an exact bulk integral. Other global aspects
of internal spinor forms and non-polynomial solutions have
been studied in \cite{BabalicFoliated,BabalicSingular,
GregoryNonpoly}.

Section~\ref{sec:review} reviews the fixed-point formula,
the GNS forms and their shift under electric monodromy.
Section~\ref{sec:general} derives the modified localization
formula and its consistency conditions.
Section~\ref{sec:physical} relates the electric period to
the magnetic flux, the Maxwell and Page charges and the
gravitational anomaly. Sections~\ref{sec:period-determination}
and \ref{sec:mixed-periodic-main} treat the compact and
periodic examples, and Section~\ref{sec:discussion}
discusses the scope and limitations of the
construction and directions for further study, including
compact mixed examples, multiple electric periods and
renormalized noncompact observables. The appendices contain the global,
spinorial and regularity derivations.
}

\section{Standard localization and electric periods}
\label{sec:review}

We first separate the fixed-point theorem from the
supergravity input that makes it applicable. The
Berline--Vergne--Atiyah--Bott formula is a global statement
about an equivariantly closed polyform. The GNS construction
supplies such forms locally from a Killing spinor, but its
global application uses an electric scalar potential.
We review both ingredients before allowing that scalar to
have nonzero periods.

\subsection{The standard equivariant fixed-point formula}
\label{sec:abbv-review}

Let $M_{2n}$ be a compact oriented Riemannian manifold
without boundary, with a Killing vector $\xi$ generating
a circle action. {Following \cite{GNS}, a polyform is a
formal sum of invariant even forms, and the equivariant
differential is}
{%
\begin{equation}\label{eq:review-polyform}
 \Phi=\sum_{j=0}^{n}\Phi_{2n-2j},
 \qquad \dd_\xi=\dd-\iota_\xi,
 \qquad \dd_\xi^2=-\mathcal L_\xi .
\end{equation}}%
The contraction $\iota_\xi$ inserts the vector $\xi$
into the first index of a differential form.
Cartan's identity
$\dd\iota_\xi+\iota_\xi\dd=\mathcal L_\xi$, together
with $\dd^2=\iota_\xi^2=0$, gives
$\dd_\xi^2=-\mathcal L_\xi$. Thus $\dd_\xi$ squares to
zero on invariant polyforms, for which
$\mathcal L_\xi\Phi=0$. The integral of a
polyform is the integral of its top component. Suppose that
{$\Phi$} is \emph{globally defined} and obeys
{$\dd_\xi\Phi=0$}. The fixed-point set is
$\operatorname{Fix}(\xi)=\{p\in M_{2n}\mid\xi(p)=0\}$.
Its connected components $\mathcal F$ are the maximal
connected subsets of this set; they can be isolated points
or positive-dimensional submanifolds. The normal bundle
$\mathcal N_{\mathcal F}$ consists of directions orthogonal
to $\mathcal F$ in $M_{2n}$, with
$TM_{2n}|_{\mathcal F}=T\mathcal F\oplus\mathcal N_{\mathcal F}$.
The standard localization theorem
\cite{BerlineVergne,AtiyahBott} then yields
{%
\begin{equation}\label{eq:review-abbv}
 \boxed{\displaystyle
 \int_{M_{2n}}\Phi
 =\sum_{\mathcal F\subset\operatorname{Fix}(\xi)}
 \frac1{d_{\mathcal F}}
 \int_{\mathcal F}
 \frac{\iota_{\mathcal F}^*\Phi}
 {e_\xi(\mathcal N_{\mathcal F})}.}
\end{equation}
Here $\iota_{\mathcal F}$ is the inclusion,
$d_{\mathcal F}$ is the order of a possible orbifold
structure group ($d_{\mathcal F}=1$ in the smooth case), and
the integral over $\mathcal F$ keeps the component of top
degree on $\mathcal F$. If the normal bundle splits into
complex lines $L_i$, on which $\xi=b_i\partial_{\varphi_i}$
with $\varphi_i\sim\varphi_i+2\pi$, the equivariant Euler
class is
\begin{equation}\label{eq:review-euler}
 e_\xi(\mathcal N_{\mathcal F})
 =\prod_{i=1}^{k}\left(
 \frac{b_i}{2\pi}+c_1(L_i)\right),
 \qquad \operatorname{codim}\mathcal F=2k,
\end{equation}
and its inverse is expanded in powers of the Chern classes.
For an isolated fixed point $p$ in eight dimensions only the
zero-form survives, and its contribution is}
\begin{equation*}
 \frac{(2\pi)^4\Phi_0(p)}
 {d_{\mathcal F}\,b_1b_2b_3b_4}.
\end{equation*}
{Thus the theorem replaces a bulk integral by fixed-locus
data.}

{%
If the circle action is free, the right-hand side of
\eqref{eq:review-abbv} is empty and the integral vanishes.
A smooth compact boundaryless $M_8$ with a free R-symmetry
and positive warped volume therefore carries no global
equivariantly closed completion of its warped-volume form.
Section~\ref{sec:general} shows that a nonzero electric
period is what prevents the local GNS completions from
gluing, and that the interfaces between electric-potential
patches then supply the volume.
}

\subsection{Supersymmetry and the local GNS forms}

We use the supersymmetry equations of \cite{Ashmore}
and the conventions of GNS \cite{GNS}.
The eleven-dimensional metric
and four-form are
\begin{align}\label{eq:ansatz}
 \dd s^2_{11}
 &=\mathrm e^{2\lambda}
  \left(\dd s_{\AdS_3}^2+\dd s_{M_8}^2\right),\nonumber\\
 G_4&=\mathrm e^{3\lambda}F
  +\vol_{\AdS_3}\wedge\mathrm e^{3\lambda}f\,.
\end{align}
The metric on $\AdS_3$ has unit radius,
and $F$ and $f$ are respectively the internal
magnetic four-form and electric one-form.
We use the source-free two-derivative equations, treating
the $C_3\wedge X_8$ anomaly contribution separately.
The Bianchi identity gives
\begin{equation}\label{eq:bianchi}
 \dd(\mathrm e^{3\lambda}F)=0,\qquad
 \dd(\mathrm e^{3\lambda}f)=0\,.
\end{equation}
For a supersymmetric bosonic background, the gravitino
variation must vanish for a nonzero spinor $\varepsilon$.
With $\Gamma_0\cdots\Gamma_{10}=-1$, the eleven-dimensional
Killing-spinor equation in our conventions is
\begin{equation}\label{eq:11d-kse}
 \delta_\varepsilon\psi_M
 =\left[\nabla_M-\frac1{288}
 \left(\Gamma_M{}^{NPQR}-8\delta_M{}^N\Gamma^{PQR}\right)
 G_{NPQR}\right]\varepsilon=0.
\end{equation}
Its reduction on $\AdS_3\times M_8$ gives the internal
differential and algebraic Killing-spinor equations of
\cite{Ashmore}. The $\mathcal N=(2,0)$ supersymmetry is
described by two Majorana spinors $\chi_1,\chi_2$ on $M_8$.
Following
\cite{GNS}, we combine them into
$\chi=(\chi_1+\mathrm i\chi_2)/\sqrt{2}$ and choose
\begin{equation}\label{eq:spinor-normalization}
 \bar\chi\chi=1,\qquad \bar\chi^c\chi=0.
\end{equation}
The eight-dimensional gamma matrices may be taken to be
real and symmetric, so $\chi^c=\chi^*$ and
$\bar\chi=\chi^\dagger$. Thus
$\bar\chi^c\chi=\chi^T\chi$, distinct from
$\bar\chi\chi=\chi^\dagger\chi$.
The chirality matrix
$\gamma_9=\gamma_1\cdots\gamma_8$ obeys $\gamma_9^2=1$.
We use scalar and differential-form bilinears of $\chi$,
\begin{align}\label{eq:bilinear-definitions}
 \zeta&=\bar\chi\gamma_9\chi,
 &S&=\bar\chi^c\gamma_9\chi,\nonumber\\
 K&=\bar\chi\gamma_{(1)}\chi,
 &\xi^\flat&=-\frac{\mathrm i}{2}
       \bar\chi\gamma_9\gamma_{(1)}\chi,\nonumber\\
 J&=-\mathrm i\bar\chi\gamma_{(2)}\chi,
 &\omega&=-\mathrm i\bar\chi\gamma_9\gamma_{(2)}\chi,
 \nonumber\\
 \Psi&=\bar\chi\gamma_{(4)}\chi.
\end{align}
Here $\gamma_{(p)}=\gamma_{\mu_1\cdots\mu_p}
\dd x^{\mu_1}\wedge\cdots\wedge\dd x^{\mu_p}/p!$,
where $\gamma_{\mu_1\cdots\mu_p}$ denotes the
antisymmetrized gamma-matrix product.
In local coordinates $x^\mu$, the two bilinear one-forms
and the metric-dual vector field are, explicitly,
\begin{equation*}
 \begin{aligned}
 K&=K_\mu\,\dd x^\mu,\\
 \xi^\flat&=\xi_\mu\,\dd x^\mu,\\
 \xi&=g^{\mu\nu}\xi_\nu\,\partial_\mu.
 \end{aligned}
\end{equation*}
The scalars $\zeta$ and $S$ are respectively real and
complex, while $K$, $\xi^\flat$, $J$, $\omega$ and $\Psi$
are real forms. The Killing-spinor
equations show that $\xi$ preserves the complete
supergravity background, and its action on the spinor is
$\mathcal L_\xi\chi=\mathrm i\chi/2$ \cite{Ashmore,GNS}.
For the special $SU(3)$ branch used later we also need
the complex one-form
\begin{equation*}
 P=\bar\chi^c\gamma_{(1)}\chi.
\end{equation*}
The bilinears made with $\bar\chi$ are invariant under
the R-symmetry flow, while $S$ and $P$, which involve
$\bar\chi^c$, have charge one,
{%
\begin{equation*}
 \mathcal L_\xi S=\mathrm iS,\qquad
 \mathcal L_\xi P=\mathrm iP.
\end{equation*}
The Killing spinor therefore fixes the R-symmetry vector
used for localization.}

A Fierz identity gives the norm of this vector,
\begin{equation}\label{eq:spinor-fixed-locus}
 \|\xi\|^2=\frac14\|K\|^2+\frac14|S|^2,
 \qquad \{\xi=0\}=\{K=0,\ S=0\}.
\end{equation}

Differentiating the bilinears and substituting the
Killing-spinor equations gives their differential
relations. For example,
$\partial_\mu\zeta=(\nabla_\mu\bar\chi)\gamma_9\chi
+\bar\chi\gamma_9\nabla_\mu\chi$. The resulting identities are
\begin{align}\label{eq:bilinear}
 \dd(\mathrm e^{3\lambda}\zeta)
 &=\mathrm e^{3\lambda}f+2\mathrm e^{3\lambda}K,
 &\dd(\mathrm e^{3\lambda}\omega)
 &=-2\iota_\xi(\mathrm e^{3\lambda}F),
 \nonumber\\
 \dd(\mathrm e^{6\lambda}\Psi)
 &=-\mathrm e^{3\lambda}K\wedge
                \mathrm e^{3\lambda}F,
 &\dd(\mathrm e^{9\lambda}*J)
 &=4\iota_\xi(\mathrm e^{9\lambda}\vol_8).
\end{align}
Here $*$ is the Hodge star for $\dd s_{M_8}^2$.
Independently, algebraic Fierz rearrangements give
\begin{equation}\label{eq:contractions}
 \iota_\xi\omega=\frac12K,\qquad
 \iota_\xi\Psi=-\frac12\omega\wedge K,\qquad
 \iota_\xi *J=-\frac12\Psi\wedge K\,.
\end{equation}

The Bianchi identity
$\dd(\mathrm e^{3\lambda}f)=0$ in \eqref{eq:bianchi}
shows that the electric one-form is closed. By the
Poincar\'e lemma, it has a scalar potential $a_0$ on
each contractible patch, although the potential need not
exist globally.
\begin{equation}\label{eq:local-electric-potential}
 \mathrm e^{3\lambda}f=\dd a_0.
\end{equation}
We then define
\begin{equation}\label{eq:y}
y=\frac12\left(\mathrm e^{3\lambda}\zeta-a_0\right),
\qquad \dd y=\mathrm e^{3\lambda}K\,.
\end{equation}
The first identity in \eqref{eq:bilinear} gives
\begin{align*}
 2\dd y
 &=\dd(\mathrm e^{3\lambda}\zeta)-\dd a_0
   =(\mathrm e^{3\lambda}f+2\mathrm e^{3\lambda}K)
    -\mathrm e^{3\lambda}f
   =2\mathrm e^{3\lambda}K,\\
 \dd(\mathrm e^{3\lambda}K)
 &=-\tfrac12\dd(\mathrm e^{3\lambda}f)=0.
\end{align*}

{%
GNS constructed equivariantly closed forms for the magnetic
flux and its dual, for calibrated M2-branes and for the
warped volume \cite{GNS}. We need three of them. The form
$\Phi^F$ completes the magnetic four-form, and its two-form
component contains the calibration form
$\mathrm e^{3\lambda}\omega$ of supersymmetric M2-branes. The
form $\Phi^\Psi$ completes $\mathrm e^{6\lambda}\Psi$, and twice
it is cohomologous to the completion of the dual flux
$\mathrm e^{6\lambda}*F$ constructed in \cite{GNS}. The form
$\Phi$ completes the warped volume. We reproduce the local
expressions in order to identify the parts that depend on the
potentially multivalued scalar $y$.}
Using \eqref{eq:bilinear}--\eqref{eq:contractions}
and $\dd y=\mathrm e^{3\lambda}K$, the GNS
polyforms are \cite{GNS}
\begin{equation}\label{eq:gns-forms}
 \boxed{\begin{aligned}
 \Phi^F&=\mathrm e^{3\lambda}F
       -\frac12\mathrm e^{3\lambda}\omega-\frac y4,\\
 \Phi^\Psi&=\mathrm e^{6\lambda}\Psi
       +y\mathrm e^{3\lambda}F
       -\frac y2\mathrm e^{3\lambda}\omega
       -\frac{y^2}{8},\\
 \Phi&=\mathrm e^{9\lambda}\vol_8
       +\frac14\mathrm e^{9\lambda}*J
       -\frac y8\mathrm e^{6\lambda}\Psi
       -\frac{y^2}{16}\mathrm e^{3\lambda}F
       +\frac{y^2}{32}\mathrm e^{3\lambda}\omega
       +\frac{y^3}{192}.
 \end{aligned}}
\end{equation}
{Each of them is a polyform in the sense of
\eqref{eq:review-polyform}.}
The coefficients of the lower-degree terms
are fixed by requiring $\dd_\xi\Phi^F=0$ degree by degree.
Starting from the closed four-form $\mathrm e^{3\lambda}F$,
the derivative of the two-form must cancel its contraction
by $\xi$. Since
$\dd(\mathrm e^{3\lambda}\omega)
=-2\iota_\xi(\mathrm e^{3\lambda}F)$, this fixes the
two-form to $-\mathrm e^{3\lambda}\omega/2$.
The remaining one-form is cancelled by the derivative of
the scalar term. The identities $\iota_\xi\omega=K/2$
and $\dd y=\mathrm e^{3\lambda}K$ fix that term to $-y/4$,
up to an additive constant. Explicitly, the five-, three-
and one-form equations are
\begin{align}\label{eq:magnetic-descent}
 \dd(\mathrm e^{3\lambda}F)&=0,\nonumber\\
 \dd\left(-\frac12\mathrm e^{3\lambda}\omega\right)
 &=\iota_\xi(\mathrm e^{3\lambda}F),
 \nonumber\\
 \dd\left(-\frac y4\right)
 &=-\frac14\mathrm e^{3\lambda}K
  =\iota_\xi\left(-\frac12\mathrm e^{3\lambda}\omega\right).
\end{align}

{%
The descents of $\Phi^\Psi$ and $\Phi$ follow in the same
way from \eqref{eq:bilinear} and \eqref{eq:contractions},
and Appendix~\ref{app:physical-pairing-proof} displays each
step. On every electric-potential patch they establish
\begin{equation}\label{eq:local-closure}
 \dd_\xi\Phi^F=\dd_\xi\Phi^\Psi
 =\dd_\xi\Phi=0.
\end{equation}
These equivariant closure equations follow from the
Killing-spinor and Fierz identities alone. Electric
monodromy affects only the gluing of the local forms
between patches.
}

\subsection{\texorpdfstring{{Electric periods and the shift of the GNS forms}}{Electric periods and the shift of the GNS forms}}
\label{sec:monodromy-review}

{%
Although $\dd y=\mathrm e^{3\lambda}K$ is global, $y$ need
not be single valued. Since $\mathrm e^{3\lambda}\zeta$ is a
global scalar, its differential has zero period, and the
first identity in \eqref{eq:bilinear} gives, around a closed
curve $\gamma$,
\begin{equation}\label{eq:shift}
 y\longrightarrow y+\delta,\qquad
 \delta=\oint_\gamma\dd y
 =-\frac12\oint_\gamma\mathrm e^{3\lambda}f .
\end{equation}
Taking de Rham classes in the same identity gives
$[\mathrm e^{3\lambda}f]=-2[\mathrm e^{3\lambda}K]$, so global
choices of $a_0$ and $y$ exist precisely when
$[\mathrm e^{3\lambda}f]=0$. GNS assume this condition
whenever they apply localization globally
\cite[eq.~(2.17)]{GNS}, and in this paper we allow it to fail.

The R-symmetry orbits carry no electric period. Indeed, the
first identity in \eqref{eq:contractions} and
$\iota_\xi^2=0$ give $\iota_\xi K=0$, and invariance of the
scalar bilinear gives
\begin{align*}
 \iota_\xi(\mathrm e^{3\lambda}f)
 &=\iota_\xi\dd(\mathrm e^{3\lambda}\zeta)
   -2\iota_\xi(\mathrm e^{3\lambda}K)\\
 &=\mathcal L_\xi(\mathrm e^{3\lambda}\zeta)=0.
\end{align*}
Integrating $\iota_\xi(\mathrm e^{3\lambda}K)=0$ along a
closed orbit then gives
}
\begin{equation}\label{eq:horizontal-electric}
 \iota_\xi(\mathrm e^{3\lambda}K)=0,\qquad
 \iota_\xi(\mathrm e^{3\lambda}f)=0,\qquad
 \oint_{\text{R-orbit}}\mathrm e^{3\lambda}K=0 .
\end{equation}
{%
The cycles with nonzero period are therefore distinct from
the R-symmetry orbits used in localization. In the compact
examples they lie along an additional circle, while the
R-symmetry acts on the three-sphere factors. When all
electric periods vanish, $a_0$ and $y$ can be chosen
globally.

All other bilinears in \eqref{eq:gns-forms} are global, so
the change of the local forms under \eqref{eq:shift}
follows by substituting $y+\delta$ for $y$. For the
warped-volume form the substitution gives
\begin{align*}
 \Phi(y+\delta)-\Phi(y)
 ={}&-\frac{\delta}{8}\mathrm e^{6\lambda}\Psi
 -\frac{2y\delta+\delta^2}{16}\mathrm e^{3\lambda}F\\
 &+\frac{2y\delta+\delta^2}{32}\mathrm e^{3\lambda}\omega
 +\frac{3y^2\delta+3y\delta^2+\delta^3}{192}\\
 ={}&-\frac{\delta}{8}\Bigl[\mathrm e^{6\lambda}\Psi
 +y\mathrm e^{3\lambda}F-\frac y2\mathrm e^{3\lambda}\omega
 -\frac{y^2}{8}\Bigr]\\
 &-\frac{\delta^2}{16}\Bigl[\mathrm e^{3\lambda}F
 -\frac12\mathrm e^{3\lambda}\omega-\frac y4\Bigr]
 +\frac{\delta^3}{192},
\end{align*}
and the two brackets are $\Phi^\Psi(y)$ and $\Phi^F(y)$. The
three shifts are therefore
}
\begin{equation}\label{eq:monodromy}
 \boxed{\begin{aligned}
 \Phi^F(y+\delta)-\Phi^F(y)
   &=-\frac{\delta}{4},\\
 \Phi^\Psi(y+\delta)-\Phi^\Psi(y)
   &=\delta\,\Phi^F(y)-\frac{\delta^2}{8},\\
 \Phi(y+\delta)-\Phi(y)
   &=-\frac{\delta}{8}\Phi^\Psi(y)
     -\frac{\delta^2}{16}\Phi^F(y)
     +\frac{\delta^3}{192}.
 \end{aligned}}
\end{equation}
{%
The polynomial shifts compose under successive electric
cycles. The dual-flux completion of \cite{GNS} differs from
$2\Phi^\Psi$ by $\dd_\xi$ of a global form, so it shifts by
$2\delta\Phi^F-\delta^2/4$. The global analysis of the next
section uses only these shifts and the local closure
\eqref{eq:local-closure}.
}

\section{Global localization with electric monodromy}
\label{sec:general}

{%
The target of this section is the leading central charge,
which is proportional to the warped volume. We construct a
local primitive of the warped volume from the GNS form
$\Phi$ and apply Stokes' theorem on the space cut open along
the interfaces of electric-potential patches.
Section~\ref{sec:modified-gns} derives the resulting
formula, Section~\ref{sec:global-consistency} shows that it
is independent of the choices involved and evaluates the cut
term cohomologically. We then obtain a corollary
constraining compact supersymmetric backgrounds without
R-symmetry fixed points.
}

\subsection{\texorpdfstring{{Volume descent across electric cuts}}{Volume descent across electric cuts}}
\label{sec:modified-gns}

{%
The leading Brown--Henneaux central charge is}
\begin{equation}\label{eq:central-charge-volume}
 c^{(0)}=\frac{3}{32\pi^7\ellp^9}
 \int_{M_8}\mathrm e^{9\lambda}\vol_8 .
\end{equation}
For compact smooth backgrounds the integral is positive.
We use the two-derivative normalization in this expression
\cite{BrownHenneaux,GNS}.

{%
We want to express this integral through boundary data
when the electric scalar $y$ is multivalued. Let $M_8$
denote the compact internal space or, for a noncompact
space, a regulated domain with outer boundary
$\partial_{\mathrm{out}}M_8$. The physical fields and the
spinor bilinears are smooth and globally defined, while
$\dd y_\alpha=\mathrm e^{3\lambda}K$ holds on
electric-potential patches $U_\alpha$. The only nontrivial
transitions are the constant shifts
$y_\alpha-y_\beta=\delta_{\alpha\beta}$. We cut $M_8$ along
interfaces $\mathcal C_{\alpha\beta}$, chosen away from the
zeros of $\xi$, into cells $M_\alpha$ on which $y_\alpha$ is
single valued, and we remove small tubular neighborhoods of
the fixed components of $\xi$.

On the patch $U_\alpha$ the warped-volume form of
\eqref{eq:gns-forms} is
\begin{equation}\label{eq:modified-gns-polyform}
 \Phi(y_\alpha)=\Phi_8+\Phi_6+\Phi_4(y_\alpha)+\Phi_2(y_\alpha)
 +\Phi_0(y_\alpha),
 \qquad \dd_\xi\Phi(y_\alpha)=0,
\end{equation}}%
{%
where $\Phi_{2k}$ is the component of form
degree $2k$. Here $\Phi_8=\mathrm e^{9\lambda}\vol_8$ and
$\Phi_6=\frac14\mathrm e^{9\lambda}*J$ are global, while
$\Phi_4$, $\Phi_2$ and $\Phi_0$ depend on $y$. The closure is
equivalent to the descent}
\begin{equation}\label{eq:volume-descent}
 \dd\Phi_{8-2j}=\iota_\xi\Phi_{10-2j},
 \qquad j=1,\ldots,4.
\end{equation}

To use Stokes' theorem, we need an ordinary
seven-form whose derivative is the warped-volume form.
The equivariantly closed $\Phi$ supplies such a primitive
where the R-symmetry vector does not vanish. We first
normalize the one-form along that vector and take its
curvature,
\begin{equation}\label{eq:eta}
 \eta=\frac{\xi^\flat}{\|\xi\|^2},\qquad
 \iota_\xi\eta=1,\qquad
 \Omega=\dd\eta,\qquad \iota_\xi\Omega=0.
\end{equation}
Here $\xi^\flat=g(\xi,\cdot)$ is a one-form,
and $\iota_\xi\eta=\eta(\xi)=1$ means that evaluating
$\eta$ on the vector $\xi$ gives one. Invariance of
$\eta$ and Cartan's identity imply
$\iota_\xi\Omega=0$, so $\Omega$ has no component along
the R-symmetry orbit. The corresponding local primitive is
\begin{equation}\label{eq:dQ}
 Q_7(y)=-\eta\wedge
 \left(\Phi_6+\Omega\wedge\Phi_4+
        \Omega^2\wedge\Phi_2+\Omega^3\Phi_0\right),
 \qquad \dd Q_7(y)=\mathrm e^{9\lambda}\vol_8.
\end{equation}
To verify the second equality, use the
descent equations \eqref{eq:volume-descent} and
$\dd\eta=\Omega$. Apart from
$\eta\wedge\iota_\xi\Phi_8=\Phi_8$, all resulting terms
are horizontal eight-forms and therefore vanish, since
the space transverse to $\xi$ has dimension seven. This
is the seven-form part of the equivariant homotopy
derived in Appendix~\ref{app:cut-pairing}.

Substituting the components of \eqref{eq:gns-forms},
\begin{align*}
 \Phi_6&=\tfrac14\mathrm e^{9\lambda}*J,
 &\Phi_4&=-\tfrac y8\mathrm e^{6\lambda}\Psi
            -\tfrac{y^2}{16}\mathrm e^{3\lambda}F,\\
 \Phi_2&=\tfrac{y^2}{32}\mathrm e^{3\lambda}\omega,
 &\Phi_0&=\tfrac{y^3}{192},
\end{align*}
into \eqref{eq:dQ} gives
\begin{align}\label{eq:Q7}
 Q_7(y)={}&-\frac14\eta\wedge \mathrm e^{9\lambda}*J
 +\eta\wedge\Omega\wedge
 \left(\frac y8\mathrm e^{6\lambda}\Psi
       +\frac{y^2}{16}\mathrm e^{3\lambda}F\right)\nonumber\\
 &-\frac{y^2}{32}\eta\wedge\Omega^2
       \wedge\mathrm e^{3\lambda}\omega
 -\frac{y^3}{192}\eta\wedge\Omega^3 .
\end{align}

{%
A nonzero electric period makes this seven-form
multivalued. Let $\mathcal C$ be an oriented cut and let
$y_-$ and $y_+$ denote the limiting values on its two faces
of the \emph{same} local primitive. For a closed curve
$\gamma$ crossing the cut once in the positive direction,
\eqref{eq:shift} gives $y_+=y_-+\delta$. Reversing $\gamma$
reverses both the sign of $\delta$ and the orientation of
the cut in Stokes' theorem.}
Consequently the local seven-form can have an additive
monodromy, while its exterior derivative remains global.
\begin{equation}\label{eq:jump-Q7}
 \Delta_\delta Q_7\equiv Q_7(y+\delta)-Q_7(y),
 \qquad \dd\Delta_\delta Q_7=0.
\end{equation}
Substitution in \eqref{eq:Q7} gives
\begin{align}\label{eq:jump-Q7-explicit}
 \Delta_\delta Q_7
 ={}&\eta\wedge\Omega\wedge
 \left[\frac{\delta}{8}\mathrm e^{6\lambda}\Psi+
       \frac{2y\delta+\delta^2}{16}\mathrm e^{3\lambda}F\right]
 \nonumber\\
 &-\frac{2y\delta+\delta^2}{32}\,
    \eta\wedge\Omega^2\wedge\mathrm e^{3\lambda}\omega
 -\frac{3y^2\delta+3y\delta^2+\delta^3}{192}
    \eta\wedge\Omega^3.
\end{align}

{%
To integrate \eqref{eq:dQ}, consider first a compact
background with no zeros of $\xi$ and one electric cycle.
The cut $\mathcal C$ is a closed seven-dimensional
hypersurface whose complement carries a single-valued
branch of $y$, and we call such cuts admissible. For an
electric one-form along $z\sim z+\mathcal L_z$, the
hypersurface $\mathcal C=\{z=z_0\}$ is admissible. For
several independent periods one uses compatible cuts or a
cell decomposition of the potential patches. Stokes'
theorem is applied to the cut-open space.}

On the cut-open manifold, let $\mathcal C_+$ and
$\mathcal C_-$ denote the two boundary copies of
$\mathcal C$.  Their induced boundary orientations are
opposite, whereas the chosen values of the local potential
obey $y_+=y_-+\delta$.  Stokes' theorem therefore reads
\begin{align*}
 \int_{M_8}\mathrm e^{9\lambda}\vol_8
 &=\int_{M_8\setminus\mathcal C}\dd Q_7\\
 &=\int_{\mathcal C_+}Q_7(y_+)
   +\int_{\mathcal C_-}Q_7(y_-)\\
 &=\int_{\mathcal C}\bigl[Q_7(y_+)-Q_7(y_-)\bigr],
\end{align*}
where the orientation of $\mathcal C$ is fixed by the
last equality. {The cut is an auxiliary boundary of the
potential branch, while $M_8$ itself has no boundary.}
\begin{equation}\label{eq:one-cut-stokes}
 \int_{M_8}\mathrm e^{9\lambda}\vol_8
 =\int_{\mathcal C}
       \bigl[Q_7(y+\delta)-Q_7(y)\bigr].
\end{equation}

The primitive $Q_7$ is defined only where
$\xi\ne0$, because $\eta$ contains $\|\xi\|^{-2}$.
If fixed points are present, we remove small neighbourhoods
of their connected components before applying Stokes'
theorem. As these neighbourhoods shrink, their boundary
integrals give the standard GNS fixed-point residues.
For a regulated noncompact space there is also an outer
boundary, whose integral is kept separately. Together
with the jumps across electric cuts, these are the three
boundary contributions of the same local $Q_7$.
\begin{equation}\label{eq:monodromy-definition}
 I_{\mathrm{fixed}}=\sum_{\mathcal F}I_{\mathcal F},
 \qquad
 I_{\mathrm{monodromy}}=
 \sum_a\int_{\mathcal C_a}
 [Q_7(y_a^+)-Q_7(y_a^-)].
\end{equation}
Here $\mathcal F$ runs over fixed components of $\xi$,
and $I_{\mathcal F}$ is precisely their standard
equivariant Euler residue. {Explicitly, in the
notation of \eqref{eq:review-abbv},}
{%
\begin{equation}\label{eq:fixed-residue}
 I_{\mathcal F}=\frac{1}{d_{\mathcal F}}
 \int_{\mathcal F}
 \frac{\iota_{\mathcal F}^*\Phi(y_{\mathcal F})}
      {e_\xi(\mathcal N_{\mathcal F})},
\end{equation}}%
Here $y_{\mathcal F}$ is the local branch of
the electric scalar near $\mathcal F$. The Euler class
and its weights are as in \eqref{eq:review-euler}.
An individual residue may
depend on $y_{\mathcal F}$, but the full fixed-point,
cut and outer-boundary sum does not.

{%
Collecting the three contributions, Stokes' theorem on the
cells $M_\alpha$ gives the modified fixed-point formula
\begin{equation}\label{eq:modified-gns-formula}
 \boxed{\begin{aligned}
 \int_{M_8}\mathrm e^{9\lambda}\vol_8
 &=I_{\mathrm{fixed}}+I_{\mathrm{monodromy}}
   +I_{\mathrm{outer}},\\
 I_{\mathrm{monodromy}}
 &=\sum_{\alpha<\beta}\int_{\mathcal C_{\alpha\beta}}
   \bigl(Q_{7,\alpha}-Q_{7,\beta}\bigr),\qquad
 I_{\mathrm{outer}}
 =\sum_\alpha\int_{\partial_{\mathrm{out}}M_\alpha}
   Q_{7,\alpha},
 \end{aligned}}
\end{equation}
where $Q_{7,\alpha}=Q_7(y_\alpha)$ and each face
$\mathcal C_{\alpha\beta}$ carries the boundary orientation
induced from $M_\alpha$, so that the cut sum agrees with
\eqref{eq:monodromy-definition}, and $I_{\mathrm{fixed}}$ is
the sum of the residues \eqref{eq:fixed-residue}. The cut term is computed entirely
from the jump \eqref{eq:jump-Q7-explicit} of the same GNS
primitive. The local potential near a fixed component
$\mathcal F$ selects the branch in its residue, while the
normal-bundle Euler class is unchanged. Intersections of
cuts with other boundary faces are handled by the oriented
cell decomposition. The
three terms depend on the branch separately;
Section~\ref{sec:global-consistency} shows that their sum
does not.

When $\delta=0$ on all cycles, $Q_7$ is global away from
the fixed set and the cut terms cancel, so for compact $M_8$
the formula \eqref{eq:modified-gns-formula} reduces to the
GNS fixed-point formula \eqref{eq:review-abbv}. If $M_8$ is
compact without boundary and $\xi$ has no zeros, both the
fixed-point and outer terms vanish and the electric cuts
supply the entire warped volume.
}

\subsection{\texorpdfstring{{Independence of choices and the cut pairing}}{Independence of choices and the cut pairing}}
\label{sec:global-consistency}

{%
Two questions remain after the Stokes construction. The
first is whether its result depends on the electric branch
and on the placement of the cuts. The second is whether the
cut integral has an intrinsic description on the quotient by
the R-symmetry circle.

For the first question, take overlapping potential patches
$U_\alpha$ on which $\dd y_\alpha=\mathrm e^{3\lambda}K$.
The differences $y_\alpha-y_\beta=\delta_{\alpha\beta}$ are
constant and satisfy
$\delta_{\alpha\beta}+\delta_{\beta\gamma}=\delta_{\alpha\gamma}$
on triple overlaps, so the transformations
\eqref{eq:monodromy} form a consistent system.
}

For a direct check, partition the cut-open space into
cells $M_\alpha$ with common faces
$\mathcal C_{\alpha\beta}$. Since
$\dd Q_{7,\alpha}=\mathrm e^{9\lambda}\vol_8$,
\begin{align*}
 \sum_\alpha\int_{M_\alpha}\mathrm e^{9\lambda}\vol_8
 &=\sum_{\alpha<\beta}
   \int_{\mathcal C_{\alpha\beta}}
      (Q_{7,\alpha}-Q_{7,\beta})
  +\sum_\alpha\int_{\partial_{\rm other}M_\alpha}
      Q_{7,\alpha}.
\end{align*}
Here $\partial_{\rm other}$ includes fixed-point tubes
and outer faces. Under $y_\alpha\mapsto y_\alpha+c_\alpha$,
let $R_{7,\alpha}=Q_{7,\alpha}'-Q_{7,\alpha}$.
Since the primitives have the same exterior derivative,
$\dd R_{7,\alpha}=0$.  The change in the entire boundary
sum is therefore
\begin{equation*}
 \sum_\alpha\int_{\partial M_\alpha}R_{7,\alpha}
 =\sum_\alpha\int_{M_\alpha}\dd R_{7,\alpha}=0.
\end{equation*}
The same argument on a common refinement proves
independence of admissible cut placement.

For the second question, consider a closed
seven-dimensional cut $\mathcal C$ on which the R-symmetry
circle acts freely. Write
$\pi:\mathcal C\to\mathcal B_6=\mathcal C/S^1_{\mathrm R}$
for its six-dimensional quotient, and let $\ell$ be the
flow period of $\xi$. The jump \eqref{eq:jump-Q7-explicit}
contains a common $\eta\wedge\Omega$ factor. After
separating the circle direction $\eta$ and curvature
$\Omega$, the remaining factor is a four-form on
$\mathcal B_6$, which we denote $g_\delta$. It is
constructed from the original spinor bilinears, not an
additional supergravity field. For an invariant form
$\alpha$, the horizontal projection
$\alpha_h=\alpha-\eta\wedge\iota_\xi\alpha$ removes its
circle component and descends to $\mathcal B_6$. The
particular quotient four-form used here is defined by
\begin{align}\label{eq:gdelta-pullback}
 \pi^*g_\delta={}&-\frac{\delta}{8}
       \bigl(\mathrm e^{6\lambda}\Psi\bigr)_h\nonumber\\
 &-\frac{2y\delta+\delta^2}{16}
  \bigl(\mathrm e^{3\lambda}F
    -\tfrac12\Omega\wedge\mathrm e^{3\lambda}\omega\bigr)_h
  \nonumber\\
 &+\frac{3y^2\delta+3y\delta^2+\delta^3}{192}\Omega^2.
\end{align}
All forms in \eqref{eq:gdelta-pullback} are
restricted to $\mathcal C$, and $\Omega=\pi^*\bar\Omega$.
The bilinear identities imply $\dd g_\delta=0$; the
componentwise cancellation is verified in
Appendix~\ref{app:cut-pairing}. Thus $[g_\delta]$ defines
a four-form cohomology class on the quotient, and the
cut contribution becomes the pairing
\begin{equation}\label{eq:cut-cohomological-pairing}
 \boxed{\begin{aligned}
 \Delta_\delta Q_7
 &=-\eta\wedge\Omega\wedge\pi^*g_\delta,\\
 I_{\mathrm{monodromy}}^{(\mathcal C)}
 &=\int_{\mathcal C}\Delta_\delta Q_7
 =-\ell\int_{\mathcal B_6}\bar\Omega\wedge g_\delta
 =-\ell^2\bigl\langle e\smile[g_\delta],
            [\mathcal B_6]\bigr\rangle,
 \qquad e=\left[\frac{\bar\Omega}{\ell}\right].
 \end{aligned}}
\end{equation}
Here $\oint_{S^1_{\mathrm R}}\eta=\ell$ follows from
$\eta(\xi)=1$; $\ell$ is the orbit parameter period,
and $e$ is the circle-bundle Euler class. For cuts with
outer edges the relative pairing also includes
$I_{\mathrm{outer}}$.
 
\noindent\emph{Corollary (compact backgrounds without
R-symmetry fixed points).} Let $M_8$ be a smooth compact
internal space without boundary, with $\mathrm e^\lambda>0$
and nowhere-vanishing $\xi$. Then the source-free
supersymmetric background necessarily has a nontrivial
electric cohomology class,
\begin{equation}\label{eq:free-positive-obstruction}
 \boxed{\displaystyle
 \xi\neq0\ \text{everywhere on compact }M_8
 \quad\Longrightarrow\quad
 [\mathrm e^{3\lambda}f]\neq0,\qquad b_1(M_8)\geq1.}
\end{equation}
Indeed, if $[\mathrm e^{3\lambda}f]=0$,
the electric scalar $y$ is globally defined. Since $\xi$
has no zeros, so are $\eta$ and $Q_7(y)$. Stokes' theorem
would then give
$0<\int_{M_8}\mathrm e^{9\lambda}\vol_8
=\int_{M_8}\dd Q_7=0$, a contradiction. Non-exact
electric flux and $b_1(M_8)\geq1$ follow. Since there are
no fixed points or outer boundaries, the complete warped
volume in \eqref{eq:modified-gns-formula} is supplied by
electric cuts. The nontrivial period cannot lie along an
R-symmetry orbit by \eqref{eq:horizontal-electric}.

\section{\texorpdfstring{{Physical content of the electric period}}{Physical content of the electric period}}
\label{sec:physical}

{%
The cut term of \eqref{eq:modified-gns-formula} is
controlled by the electric period $\delta$. This section
relates $\delta$ and the cut term to physical data, namely
the magnetic flux in Section~\ref{sec:general-transgression},
the electric Maxwell and Page charges in
Section~\ref{sec:free-electric-charge}, and the
gravitational anomaly together with the global completion
of the magnetic flux in
Section~\ref{sec:magnetic-obstruction-anomaly}.
}

\subsection{Magnetic flux and the electric period}
\label{sec:general-transgression}

{%
Can the electric period be reconstructed from the magnetic
flux and the R-symmetry vector alone? This matters for the
compact examples, where the period is computed from
quantized magnetic flux without using the electric field.
The magnetic descent \eqref{eq:magnetic-descent} and the
first contraction in \eqref{eq:contractions} give
\begin{equation}\label{eq:magnetic-transgression}
 \boxed{\displaystyle
 \dd(\mathrm e^{3\lambda}\omega)
 =-2\iota_\xi(\mathrm e^{3\lambda}F),\qquad
 \delta_\gamma=2\oint_\gamma
             \iota_\xi(\mathrm e^{3\lambda}\omega).}
\end{equation}
The period is thus a transgression of the magnetic flux.
The first equation determines $\mathrm e^{3\lambda}\omega$
from the magnetic four-form, and the second contracts it
once more with $\xi$.

The first equation fixes $\mathrm e^{3\lambda}\omega$ only
up to an invariant closed two-form $\vartheta_2$. Replacing
$\mathrm e^{3\lambda}\omega$ by
$\mathrm e^{3\lambda}\omega+\vartheta_2$ at fixed magnetic class
and R-symmetry vector shifts the period by}
\begin{equation}\label{eq:magnetic-electric-class}
 \delta'_\gamma-\delta_\gamma
 =2\oint_\gamma\iota_\xi\vartheta_2,
 \qquad \dd\vartheta_2=0,\qquad\mathcal L_\xi\vartheta_2=0.
\end{equation}
{%
The magnetic class and $\xi$ therefore determine the period
whenever all permitted $\iota_\xi\vartheta_2$ have zero periods,
which holds in both compact examples below. In a complete
supersymmetric solution the Killing spinor fixes $\omega$,
and $\vartheta_2$ parametrizes the freedom in reconstructing
$\omega$ from the magnetic class alone.
Appendix~\ref{app:magnetic-completion} relates this freedom
to changes of the magnetic representative, and
Section~\ref{sec:magnetic-obstruction-anomaly} uses it to
construct ordinary completions of the magnetic flux.
}

\subsection{\texorpdfstring{{The cut term and the Maxwell charge}}{The cut term and the Maxwell charge}}
\label{sec:free-electric-charge}

{%
The cut formula computes a geometric volume. We now ask
when this volume can be written through the physical
electric charge. For a compact background with a free
R-symmetry and one independent electric period, the complete
cut integral reduces to a single term linear in $\delta$.
An additional $SU(3)$ spinor condition then turns the
surviving seven-form into the Maxwell flux density. Only
this second step uses the special spinor branch.
}

More precisely, suppose $M_8$ is compact without boundary,
$\xi$ acts freely and the
periods of $\mathrm e^{3\lambda}K$ form
$\delta\mathbb Z$ with $\delta\ne0$.
{Then $y\bmod\delta$ is a smooth circle-valued
function. Choose as the cut $\mathcal C$ the complete preimage
of one of its regular values, a closed oriented hypersurface
that may be disconnected, and let $y_0$ be the value of the
potential on one face. Cutting along all components gives a
fundamental domain for the potential branch.} Substituting
\eqref{eq:jump-Q7-explicit} into
\eqref{eq:one-cut-stokes} gives
\begin{align*}
 \int_{M_8}\mathrm e^{9\lambda}\vol_8
 ={}&\frac{\delta}{8}
 \int_{\mathcal C}\eta\wedge\Omega\wedge
                     \mathrm e^{6\lambda}\Psi\\
 &+\frac{2y_0\delta+\delta^2}{16}
 \int_{\mathcal C}\eta\wedge\Omega\wedge
                \bigl(\mathrm e^{3\lambda}F
                  -\tfrac12\Omega\wedge
                   \mathrm e^{3\lambda}\omega\bigr)\\
 &-\frac{3y_0^2\delta+3y_0\delta^2+\delta^3}{192}
 \int_{\mathcal C}\eta\wedge\Omega^3.
\end{align*}
{%
The constant $y_0$ contains the arbitrary origin of the
electric potential. Shifting the integration constant in
$a_0$ shifts $y_0$ without changing the cut or the forms in
the three integrals, and by Section~\ref{sec:global-consistency}
the volume is unchanged. The right-hand side is therefore a
polynomial in $y_0$ with fixed coefficients and constant
value, so its $y_0^2$ and $y_0$ coefficients vanish.}
The first gives
\begin{equation*}
 \int_{\mathcal C}\eta\wedge\Omega^3=0.
\end{equation*}
The second gives
\begin{equation*}
 \int_{\mathcal C}\eta\wedge\Omega\wedge
                \bigl(\mathrm e^{3\lambda}F
                    -\tfrac12\Omega\wedge
                     \mathrm e^{3\lambda}\omega\bigr)=0.
\end{equation*}
{Thus the integrated higher-order jump terms
vanish under these compact rank-one assumptions, and}
\begin{equation}\label{eq:rank-one-volume}
 \int_{M_8}\mathrm e^{9\lambda}\vol_8
 =\frac{\delta}{8}\int_{\mathcal C}
 \eta\wedge\dd\eta\wedge\mathrm e^{6\lambda}\Psi.
\end{equation}
{%
The cancellation uses the closed level-set cut of a compact
space. The noncompact example of
Section~\ref{sec:mixed-periodic-main} has a cut with an edge
on the outer boundary and retains a cubic term.
}

{The surviving seven-form is geometric.} To identify it with the
electric Maxwell flux, we now impose the additional
$SU(3)$ spinor branch $\zeta=S=P=0$. The algebraic and
differential Killing-spinor equations, together with Fierz
identities, give the two relations
\begin{align}\label{eq:maxwell-two-kse-steps}
 J\mathbin{\lrcorner}\iota_\xi F+f+3K&=0,\nonumber\\
 \eta\wedge\dd\eta\wedge\Psi
 &=-4*_{\rm GNS}\bigl(J\mathbin{\lrcorner}\iota_\xi F+3K\bigr).
\end{align}
{Substituting the first identity into the second
eliminates the magnetic contraction,}
{%
\begin{equation*}
 \eta\wedge\dd\eta\wedge\Psi
 =-4*_{\rm GNS}\bigl((-f-3K)+3K\bigr)
 =4*_{\rm GNS}f,
\end{equation*}
and multiplication by $\mathrm e^{6\lambda}$ gives the pointwise relation}
\begin{equation}\label{eq:maxwell-bilinear-pairing}
 \eta\wedge\dd\eta\wedge\mathrm e^{6\lambda}\Psi
 =4\mathrm e^{6\lambda}*_{\rm GNS} f.
\end{equation}
Here $*_{\rm GNS}$ is the GNS Hodge convention,
opposite to the conventional left star on one-forms;
the full spinor-frame derivation and orientation check
are in Appendix~\ref{app:maxwell-spinor-proof}.
For the positively oriented electric charge cycle
$\Sigma_Q=-\mathcal C$, define
\begin{equation}\label{eq:maxwell-charge}
 \mathcal Q_{\rm Maxwell}=
 \frac{1}{(2\pi\ell_p)^6}
 \int_{\Sigma_Q}\mathrm e^{6\lambda}*_{\rm GNS}f.
\end{equation}
Consequently \eqref{eq:rank-one-volume} and
\eqref{eq:maxwell-bilinear-pairing} imply
{%
\begin{equation*}
 \int_{M_8}\mathrm e^{9\lambda}\vol_8
 =\frac{\delta}{2}\int_{\mathcal C}\mathrm e^{6\lambda}*_{\rm GNS}f
 =-\frac{\delta}{2}(2\pi\ell_p)^6\mathcal Q_{\rm Maxwell}.
\end{equation*}}%
The Brown--Henneaux normalization then gives
\begin{equation}\label{eq:rank-one-central-charge}
 \boxed{\displaystyle
 c^{(0)}=-\frac{3\delta}{\pi\ell_p^3}\,
 \mathcal Q_{\rm Maxwell}.}
\end{equation}
The Maxwell flux is not generally closed because the
eleven-dimensional equation of motion contains the
Chern--Simons interaction. In the present source-free,
two-derivative convention its internal component obeys
\begin{equation}
 \dd(\mathrm e^{6\lambda}*_{\rm GNS}f)
 =-\frac12\mathrm e^{3\lambda}F\wedge
                 \mathrm e^{3\lambda}F.
\end{equation}
To obtain an on-shell closed electric charge density,
one adds a magnetic-potential term. This distinguishes
Maxwell and Page charges \cite{Marolf2000}. For a specified
magnetic potential representative on $\Sigma_Q$, the
classical Page form is
\begin{equation}\label{eq:page-current}
 \mathcal J_7=\mathrm e^{6\lambda}*_{\rm GNS}f
 +\frac12 C_3^{\rm mag}\wedge\mathrm e^{3\lambda}F,
 \qquad \dd C_3^{\rm mag}=\mathrm e^{3\lambda}F.
\end{equation}
It obeys $\dd\mathcal J_7=0$ wherever the magnetic
potential is defined.
When $C_3^{\rm mag}$ admits one representative on the
electric charge cycle $\Sigma_Q$, the classical Page
charge in this polarization is defined by
\begin{equation*}
 N_2^{\rm Page}:=\frac{1}{(2\pi\ell_p)^6}
 \int_{\Sigma_Q}\mathcal J_7
 =\mathcal Q_{\rm Maxwell}
 +\frac{1}{2(2\pi\ell_p)^6}
 \int_{\Sigma_Q}C_3^{\rm mag}\wedge
                      \mathrm e^{3\lambda}F.
\end{equation*}
Thus the relation to the Page charge is
\begin{equation}\label{eq:rank-one-page}
 c^{(0)}=-\frac{3\delta}{\pi\ell_p^3}
 \left[N_2^{\rm Page}
 -\frac{1}{2(2\pi\ell_p)^6}
 \int_{\Sigma_Q}C_3^{\rm mag}\wedge
                          \mathrm e^{3\lambda}F\right].
\end{equation}
{%
These ordinary integrals require a magnetic potential
$C_3^{\rm mag}$ defined over all of $\Sigma_Q$. For
patchwise magnetic potentials, $N_2^{\rm Page}$ is a
relative Page pairing with the prescribed magnetic patching
and large-gauge choice, and the correction in
\eqref{eq:rank-one-page} then includes the associated overlap
and corner terms. In both compact examples
$\left.\mathrm e^{3\lambda}F\right|_{\Sigma_Q}=0$, so the
correction vanishes and the Maxwell and Page charges
coincide. The relation \eqref{eq:rank-one-central-charge}
holds under the compact rank-one and $SU(3)$ assumptions
stated above. The next subsection asks what the same
electric class implies for the magnetic flux.
}

\subsection{\texorpdfstring{{Gravitational anomaly and the magnetic obstruction}}{Gravitational anomaly and the magnetic obstruction}}
\label{sec:magnetic-obstruction-anomaly}

Can the electric monodromy be seen in a parity-odd observable
of the dual CFT? The $C_3\wedge X_8$ contribution to its
gravitational anomaly and its fixed-point localization
were already treated by Gregory, Notman and Sparks (GNS)
\cite{GNS}; a further analysis of trial anomalies in
projective-bundle backgrounds appears in
\cite{GregoryAnomaly}. Here we ask how that established
contribution depends on the electric class when the
R-symmetry circle has no fixed points.

The eleven-dimensional $C_3\wedge X_8$ coupling induces a
three-dimensional gravitational Chern--Simons term and
thereby contributes to the left--right central-charge
difference. In the conventions of GNS,
\begin{equation}\label{eq:anomaly}
 \left.(c_L-c_R)\right|_{C_3X_8}
 =-\frac{1}{4(2\pi\ell_p)^3}
 \int_{M_8}\mathrm e^{3\lambda}F\wedge p_1(TM_8).
\end{equation}

This is only the indicated higher-derivative contribution,
not a claim to determine every possible contribution to the
quantum gravitational anomaly.

To relate \eqref{eq:anomaly} to the electric flux, assume
that $M_8$ is smooth, compact and without boundary, and
that the R-symmetry generates a \emph{free} circle action
with flow period $\ell_\xi$. The quotient
$\pi:M_8\to B_7=M_8/U(1)_{\mathrm R}$ is then a smooth
seven-manifold. Write $\eta=\xi^\flat/\|\xi\|^2$,
$\Omega=\dd\eta$ and
$e=[\bar\Omega/\ell_\xi]\in H^2(B_7;\mathbb R)$ for
the Euler class of this circle bundle. An overbar here
denotes a form on the quotient, not complex conjugation:
for an invariant horizontal form $\alpha$ on $M_8$,
$\alpha=\pi^*\bar\alpha$. Thus
$\Omega=\pi^*\bar\Omega$ and
$\mathrm e^{3\lambda}f
=\pi^*\overline{\mathrm e^{3\lambda}f}$.
In contrast, the magnetic four-form need not be
horizontal; $\pi_*[\mathrm e^{3\lambda}F]$ denotes its
three-form class after integration along the R-symmetry
circle.

The Killing-spinor identities
\eqref{eq:magnetic-descent} determine this circle integral
without requiring an explicit magnetic potential. The
result, derived in Appendix~\ref{app:magnetic-completion},
is
\begin{equation}\label{eq:gysin-pushforward}
 \pi_*[\mathrm e^{3\lambda}F]
 =-\frac{\ell_\xi^2}{4}
 e\wedge[\overline{\mathrm e^{3\lambda}K}]
 =\frac{\ell_\xi^2}{8}e\wedge
 [\overline{\mathrm e^{3\lambda}f}].
\end{equation}

Thus the magnetic flux around the R-symmetry circle is
controlled jointly by the electric cohomology class and
the topology of that circle bundle. A nonzero electric
period alone does not guarantee a nonzero pushforward:
the cup product with $e$ may vanish.

This distinction also decides whether the magnetic
four-form possesses \emph{some} global ordinary equivariant
completion, after allowing the invariant closed two-form
freedom $\vartheta_2$ of
Section~\ref{sec:general-transgression}. The Gysin sequence
for the free action gives the criterion
\begin{equation}\label{eq:gysin-obstruction}
 \pi_*[\mathrm e^{3\lambda}F]=0
 \quad\text{in }H^3(B_7;\mathbb R).
\end{equation}

Equation \eqref{eq:gysin-obstruction} is necessary and
sufficient for such a completion to exist; its derivation,
including the $\vartheta_2$ freedom, is in
Appendix~\ref{app:magnetic-completion}. This is a
\emph{magnetic}-completion question, distinct from the
warped-volume obstruction
\eqref{eq:free-positive-obstruction}.

Since $p_1(TM_8)=\pi^*p_1(TB_7)$ in cohomology, the
existing GNS anomaly integral \eqref{eq:anomaly} can now
be written entirely in terms of forms on $B_7$:
\begin{equation}\label{eq:anomaly-electric-class}
 \boxed{\displaystyle
 \left.(c_L-c_R)\right|_{C_3X_8}
 =-\frac{\ell_\xi^2}{32(2\pi\ell_p)^3}
 \int_{B_7}e\wedge\overline{\mathrm e^{3\lambda}f}
 \wedge p_1(TB_7).}
\end{equation}

This is a free-circle reformulation of the known
$C_3\wedge X_8$ gravitational-anomaly contribution. The
additional information is that the same
$e\wedge[\overline{\mathrm e^{3\lambda}f}]$
also governs the existence of a global magnetic
completion.

\noindent\emph{Corollary (a sufficient anomaly test).}
Under the compact free-circle assumptions above,

\begin{equation}\label{eq:anomaly-test}
 \begin{gathered}
 \left.(c_L-c_R)\right|_{C_3X_8}\neq0
 \ \Longrightarrow\
 \pi_*[\mathrm e^{3\lambda}F]\neq0\\
 \Longrightarrow\
 \text{no global ordinary equivariant magnetic completion}.
 \end{gathered}
\end{equation}

The converse is false: a nonzero magnetic obstruction can
pair trivially with $p_1(TB_7)$ and give no contribution to
\eqref{eq:anomaly}. The two compact examples below
illustrate both possibilities. On
$S^3_+\times S^3_-\times T^2$ the electric period and
magnetic obstruction are nonzero, but $p_1(TM_8)=0$, so
this anomaly contribution vanishes. On
$S^3\times K3\times S^1$ the Pontryagin pairing is nonzero
and the same established anomaly formula gives
$\left.(c_L-c_R)\right|_{C_3X_8}=-12N_5$
[cf.~\eqref{eq:k3-anomaly-result}].

The implication above requires a compact space and a free
R-symmetry action; it cannot be applied unchanged at fixed
points or outer boundaries. Conversely, our electric-cut
localization formula \eqref{eq:modified-gns-formula}
does not require a globally completed magnetic form.

\section{\texorpdfstring{{Compact examples}}{Compact examples}}
\label{sec:period-anomaly}\label{sec:period-determination}

{%
Both compact solutions below are products with an electric
circle on which the R-symmetry acts trivially, and the
R-symmetry acts freely on the remaining factors.
Section~\ref{sec:spheres} computes the warped volume of
$S^3\times S^3\times T^2$ entirely from the electric cut.
Section~\ref{sec:physical-R-symmetry} recovers the same
electric period from the quantized magnetic flux and
compares the resulting polynomial in the trial R-symmetry
with the bulk R-current levels. Section~\ref{sec:k3} treats
$S^3\times K3\times S^1$, where the $C_3\wedge X_8$
contribution is nonzero and detects the magnetic
obstruction.
}

\subsection{\texorpdfstring{{The $S^3_+\times S^3_-\times T^2$ solution and its electric cut}}{The S3 x S3 x T2 solution and its electric cut}}
\label{sec:spheres}

Let $z,w$ have periods $\mathcal L_z,\mathcal L_w$. The
$S^3_+\times S^3_-\times T^2$ solution
\cite{Boonstra1998,Gukov2004}
(also recovered as the constant-Toda solution
in the large $\mathcal N=(4,4)$ classification
\cite{ContiMacpherson2024}) is
\begin{align}\label{eq:spheres-fields}
 \dd s^2_{11}={}&L^2\dd s^2_{\AdS_3}
      +R_+^2\dd s^2_{S^3_+}+R_-^2\dd s^2_{S^3_-}
      +\dd z^2+\dd w^2,\nonumber\\
 G_4={}&\left(\frac2L\vol_{\AdS_3}
      +\frac2{R_+}\vol_{S^3_+}
      +\frac2{R_-}\vol_{S^3_-}\right)\wedge\dd z,
 \qquad \frac1{L^2}=\frac1{R_+^2}+\frac1{R_-^2}.
\end{align}
The magnetic flux threads $S^3_\pm\times S^1_z$, the electric
component has its internal leg along $z$, and
$\mathrm e^\lambda=L$. All volume forms carry physical radii. The radius relation
follows from the eleven-dimensional Einstein equations
and is compatible with the Killing-spinor projectors.

We choose the $\mathcal N=(2,0)$ Killing spinor obtained by
selecting one complex supercharge of the enhanced supersymmetry.
Its bilinears have
\begin{equation}\label{eq:spheres-basic}
 \zeta=0,\qquad K=-\frac{\dd z}{L},\qquad
 \|\xi\|^2=\frac14,\qquad \Psi=-\frac12J\wedge J.
\end{equation}
The vector $\xi$ is a linear combination of the two Hopf
generators on the three-spheres. Both generators have constant
nonzero norm and the two sphere factors are orthogonal.
Consequently $\xi$ is everywhere nonvanishing. The full Killing-spinor
projector and the differential-form identities are checked in
Appendix~\ref{app:spinors}. The supersymmetric
spinor fixes the coefficients of the Hopf generators,
as required by the physical R-symmetry.

The electric one-form in GNS conventions is
\begin{equation}\label{eq:spheres-period}
 \mathrm e^{3\lambda}f=2L^2\dd z,
 \qquad \dd y=-L^2\dd z,
 \qquad \delta=-L^2\mathcal L_z.
\end{equation}
Only $S^1_z$ has a nonzero electric period; Hopf circles are
contractible inside the three-spheres. Thus $y$ changes by
$\delta$ around $S^1_z$, although the metric, flux and spinor
remain globally regular.

{%
The fixed-point set of $\xi$ is empty and $M_8$ has no
boundary, so \eqref{eq:modified-gns-formula} states that the
entire warped volume is the cut term. We evaluate it before
using any magnetic transgression or Chern--Simons level.
}

Cut the $z$ circle at one point. The cut has the topology
$S^3_+\times S^3_-\times S^1_w$ and is transverse to
the electric direction. On this cut the magnetic flux and
two-form $\omega$ pull back to zero, and $(\dd\eta)^3=0$.
Equation~\eqref{eq:Q7} therefore reduces to
\begin{equation}\label{eq:spheres-jump}
 \left.\bigl[Q_7(y+\delta)-Q_7(y)\bigr]\right|_{\mathcal C}
 =\frac\delta8\,\eta\wedge\dd\eta
               \wedge\mathrm e^{6\lambda}\Psi\,.
\end{equation}
Indeed, $\mathrm e^{3\lambda}F$ and
$\mathrm e^{3\lambda}\omega$ each contain $\dd z$,
while $\dd\eta$ is the sum of two Hopf curvatures on
two-dimensional horizontal spaces. The spinor bilinears give
\begin{equation}\label{eq:spheres-local}
 \eta\wedge\dd\eta\wedge\Psi
 =-8\,\vol_{S^3_+\times S^3_-\times S^1_w}
       \big|_{\dd s_{M_8}^2},
\end{equation}
with the orientation inherited from the full eight-manifold.
{The wedge multiplication in Appendix~\ref{app:spinors}
uses the spinor-selected Hopf vector and the radius relation.}
{The internal metric in \eqref{eq:ansatz} is the physical
metric divided by $L^2$, so the cut volume is}
{%
\begin{equation*}
 \int_{\mathcal C}\vol_{S^3_+\times S^3_-\times S^1_w}
 \big|_{\dd s_{M_8}^2}
 =\Bigl(2\pi^2\frac{R_+^3}{L^3}\Bigr)
  \Bigl(2\pi^2\frac{R_-^3}{L^3}\Bigr)\frac{\mathcal L_w}{L}
 =\frac{4\pi^4R_+^3R_-^3\mathcal L_w}{L^7}.
\end{equation*}}%
Substituting the negative electric period and
\eqref{eq:spheres-local} into the jump gives
\begin{align}\label{eq:spheres-jump-integration}
 \int_{\mathcal C}\bigl[Q_7(y+\delta)-Q_7(y)\bigr]
 &=-\delta L^6\,
      \frac{4\pi^4R_+^3R_-^3\mathcal L_w}{L^7}
      \nonumber\\
 &=4\pi^4 L R_+^3R_-^3
           \mathcal L_z\mathcal L_w.
\end{align}
The fixed-point contribution vanishes, so the
cut calculation yields
\begin{equation}\label{eq:spheres-volume}
 \int_{M_8}\mathrm e^{9\lambda}\vol_8
 =4\pi^4 L R_+^3R_-^3\mathcal L_z\mathcal L_w\,.
\end{equation}
{The result equals the direct volume integral. Its
factor of $L$ follows from the GNS metric normalization,
since the GNS internal metric is $g_{M_8}=L^{-2}g_{8,\rm phys}$,}
so $\vol_8=L^{-8}\vol_{8,\rm phys}$ and
\begin{equation}\label{eq:spheres-volume-rescaling}
 \mathrm e^{9\lambda}\vol_8
 =L^9L^{-8}\vol_{8,\rm phys}
 =L\,\vol_{8,\rm phys}.
\end{equation}

The magnetic flux integrals, with physical sphere volumes
$2\pi^2R_\pm^3$, give
\begin{align}\label{eq:spheres-magnetic-flux}
 N_5^\pm
 &=\frac{1}{(2\pi\ellp)^3}
   \int_{S^3_\pm\times S^1_z}G_4\nonumber\\
 &=\frac{1}{(2\pi\ellp)^3}
   \frac{2}{R_\pm}(2\pi^2R_\pm^3)\mathcal L_z
  =\frac{R_\pm^2\mathcal L_z}{2\pi\ellp^3}.
\end{align}
Combining this magnetic quantization with the radius
relation in \eqref{eq:spheres-fields} rewrites the directly
measured electric period as
\begin{equation}\label{eq:spheres-direct-charge-period}
 \delta_{\mathrm{direct}}=-L^2\mathcal L_z
 =-2\pi\ell_p^3\frac{N_5^+N_5^-}{N_5^++N_5^-}.
\end{equation}
This expression will be tested independently by magnetic
transgression below.
The dual electric flux on $S^3_+\times S^3_-\times S^1_w$ gives
\begin{align}\label{eq:spheres-electric-flux}
 N_2
 &=\frac{1}{(2\pi\ellp)^6}
     \int_{S^3_+\times S^3_-\times S^1_w}*_{11}G_4
  \nonumber\\
 &=\frac{2}{L(2\pi\ellp)^6}
    (2\pi^2R_+^3)(2\pi^2R_-^3)\mathcal L_w
  =\frac{R_+^3R_-^3\mathcal L_w}{8\pi^2L\ellp^6}.
\end{align}
We choose positive brane-charge orientations \cite{GNS};
$N_2$ is the two-derivative charge, before higher-derivative
Page corrections.
Substitution of \eqref{eq:spheres-magnetic-flux}
and \eqref{eq:spheres-electric-flux}, together with
the radius relation in \eqref{eq:spheres-fields}
into the Brown--Henneaux expression then gives
the leading central charge
\begin{equation}\label{eq:spheres-central}
 \boxed{\displaystyle
 c^{(0)}=\frac{6N_2N_5^+N_5^-}{N_5^++N_5^-}.}
\end{equation}
{%
The entire warped volume, and hence the central charge,
comes from the electric cut. The same value follows from the
Maxwell form \eqref{eq:rank-one-central-charge} of the cut
term. The $SU(3)$ identity \eqref{eq:maxwell-bilinear-pairing}
applies, and the magnetic correction in \eqref{eq:rank-one-page}
vanishes because $\mathrm e^{3\lambda}F$ restricts to zero on
the charge cycle, so
\begin{equation}\label{eq:period-page-pairing}
 \int_{\mathcal C}\eta\wedge\dd\eta\wedge
 \mathrm e^{6\lambda}\Psi=-4(2\pi\ell_p)^6N_2 .
\end{equation}
With $\delta=\delta_{\rm direct}$, the relation
$c^{(0)}=-3\delta N_2/(\pi\ell_p^3)$ reproduces
\eqref{eq:spheres-central}.
}

\subsection{\texorpdfstring{{Electric monodromy and R-current levels}}{Electric monodromy and R-current levels}}
\label{sec:physical-R-symmetry}

Section~\ref{sec:spheres} obtained the electric period by
integrating the explicit electric one-form. We now
reconstruct it from the \emph{quantized magnetic flux} and
the Killing-spinor identities, without using that electric
integral. We then compare the result with an independent
three-dimensional current-algebra calculation. The
comparison is useful because the two calculations use
different parts of eleven-dimensional supergravity.

Each three-sphere has a Hopf-circle rotation, generated by
$H_+$ or $H_-$. We normalize these generators to have
period $2\pi$ and to act on the selected supercharge with
charge $+1$. The dual CFT's unit-charge trial R-symmetry is
therefore the mixture
$R_t=tH_++(1-t)H_-$; the coefficients add to one so the
supercharge retains unit R-charge. The Killing spinor
determines which value of $t$ is supersymmetric.

The spinor bilinear $\xi^\flat=g_{M_8}(\xi,\cdot)$ is a
one-form, and raising its index with the internal metric
gives the R-symmetry vector. With the sphere Hopf one-forms
$\eta_\pm$, the result is
\begin{equation}\label{eq:spinor-hopf-generator}
 \xi^\flat=\frac12(\eta_++\eta_-),\qquad
 R=2\xi=\frac{L^2}{R_+^2}H_+
          +\frac{L^2}{R_-^2}H_- .
\end{equation}

The full spinor computation and Hopf normalization are in
Appendix~\ref{app:spinors}. Using the radius relation
\eqref{eq:spheres-fields} and the magnetic charges
\eqref{eq:spheres-magnetic-flux}, the spinor selects
$t_{\rm BPS}=L^2/R_+^2
=N_5^-/(N_5^++N_5^-)$.

The magnetic descent uses the trial generator twice:
first in
$\dd(\mathrm e^{3\lambda}\omega)
=-2\iota_\xi(\mathrm e^{3\lambda}F)$
and then in
$\mathrm e^{3\lambda}K
=2\iota_\xi(\mathrm e^{3\lambda}\omega)$.
This explains why its period is quadratic in the Hopf
weights. Applying the normalized Hopf transgression to the
two magnetic flux classes gives
\begin{equation}\label{eq:period-trial}
 \delta_{\rm tr}(t)=-2\pi\ell_p^3
 \left[N_5^+t^2+N_5^-(1-t)^2\right].
\end{equation}

The derivation, including the fact that the invariant
closed two-form ambiguity cannot change the electric
$S^1_z$ period in this geometry, is in
Appendix~\ref{app:magnetic-completion}. For general $t$,
$\delta_{\rm tr}(t)$ is a trial transgression polynomial;
only at the spinor-selected value is it identified with
the electric period of this supersymmetric background.
Thus
\begin{equation}\label{eq:trial-versus-physical}
 \boxed{\displaystyle
 \delta=\delta_{\rm tr}(t_{\rm BPS})
 =-2\pi\ell_p^3\frac{N_5^+N_5^-}{N_5^++N_5^-}.}
\end{equation}

This reproduces the direct electric-field result
\eqref{eq:spheres-direct-charge-period} from the magnetic
charges and the spinor-selected R-symmetry.

We next ask whether the same quadratic dependence appears
in the boundary currents. Gauging the
$SU(2)_L\times SU(2)_R$ isometries of each three-sphere
produces three-dimensional gauge fields. Their
Chern--Simons couplings fix the levels of the
corresponding two-dimensional current algebras. The
background has $G_4\wedge G_4=0$, so a background
$C_3\wedge G_4\wedge G_4$ term does not by itself supply
these levels. In the electric \emph{Page-fixed} ensemble,
gauging the sphere magnetic flux instead produces the
coupling through the electric kinetic term and its
boundary Legendre transform. The reduction is given in
Appendix~\ref{app:bulk-current}.

In the standard $SU(2)$ normalization the two sphere
current algebras have levels $k_\pm=N_2N_5^\pm$.
The Hopf generators are twice the Cartan generators,
so their Abelian levels are
\begin{equation}\label{eq:bulk-levels}
 k_\pm=N_2N_5^\pm,\qquad
 k_{H_\pm H_\pm}=2N_2N_5^\pm,\qquad
 k_{H_+H_-}=0.
\end{equation}

These sphere-isometry levels are consistent with the
established results
\cite{Gukov2004,HansenKraus,HST2025}. The Page-fixed
eleven-dimensional calculation fixes the normalization
used here, independently of the magnetic transgression.
For the trial mixture $R_t$ the two calculations yield
\begin{equation}\label{eq:current-trial}
 \boxed{\displaystyle
 k_{R_tR_t}^{\rm bulk}
 =2N_2\left[N_5^+t^2+N_5^-(1-t)^2\right]
 =-\frac{N_2}{\pi\ell_p^3}\delta_{\rm tr}(t).}
\end{equation}

The R-current level and the reconstructed electric
period therefore encode the same dependence on the
magnetic charges, although they were obtained from
different supergravity calculations. Their common
stationary point is
\begin{equation}\label{eq:physical-r-weight}
 t_*=\frac{N_5^-}{N_5^++N_5^-},\qquad
 \left.\partial_t\delta_{\rm tr}(t)\right|_{t_*}=0.
\end{equation}

It coincides with $t_{\rm BPS}$ fixed by the Killing
spinor. Equivalently, current-algebra orthogonality or
$c$-extremization \cite{BeniniBobevExact} selects the same
R-symmetry. Finally,
$3k_{R_{t_*}R_{t_*}}
=6N_2N_5^+N_5^-/(N_5^++N_5^-)$,
precisely the leading central charge
\eqref{eq:spheres-central}. The agreement links the
spinor-selected R-symmetry, magnetic reconstruction of
electric monodromy and Page-fixed bulk current levels.

\subsection{\texorpdfstring{{Gravitational anomaly on $S^3\times K3\times S^1$}}{Gravitational anomaly on S3 x K3 x S1}}
\label{sec:k3}

The first compact background has electric monodromy and a
magnetic-completion obstruction, but its
$C_3\wedge X_8$ gravitational-anomaly contribution
vanishes because $p_1(TM_8)=0$. Replacing one
three-sphere and one circle by K3 provides a complementary
test: the electric monodromy remains, while the nonzero K3
curvature allows the established anomaly formula
\eqref{eq:anomaly} to detect the obstruction.

The background is a constant-warp-factor
specialization of the Calabi--Yau two-fold solutions of
\cite{LozanoMacphersonNunezRamirez2019} and their
eleven-dimensional uplift \cite{LozanoNunezRamirezSpeziali2020}.
Let K3 carry any smooth hyper-K\"ahler metric of volume
$V_{K3}$. The supersymmetric product solution is
\begin{equation}\label{eq:k3-fields}
 \dd s^2_{11}=L^2\left(\dd s^2_{\AdS_3}
                      +\dd s^2_{S^3}\right)
               +\dd s^2_{K3}+\dd z^2,
 \qquad
 G_4=\frac2L\left(\vol_{\AdS_3}+\vol_{S^3}\right)
                      \wedge\dd z,
\end{equation}

where $z\sim z+\mathcal L_z$ and the volume forms
include their physical radii. The Killing spinor is
parallel along K3, and the remaining bilinears agree with
the one-sphere reduction of Section~\ref{sec:spheres}
(Appendix~\ref{app:spinors}). The R-symmetry acts along
the Hopf circle of $S^3$ without fixed points. The
electric one-form and its nonzero period are
\begin{equation}\label{eq:k3-period}
 K=-\frac{\dd z}{L},\qquad
 \|\xi\|^2=\frac14,\qquad
 \mathrm e^{3\lambda}f=2L^2\dd z,
 \qquad \delta=-L^2\mathcal L_z .
\end{equation}

Since the internal space is compact and the R-symmetry
has no fixed points, the warped volume is supplied entirely
by a cut across the electric circle. The cut is
$S^3\times K3$, and its jump is linear in $\delta$,
as in \eqref{eq:spheres-jump}. The explicit spinor-form
wedge multiplication is given in
Appendix~\ref{app:k3-curvature}. It yields
\begin{equation}\label{eq:k3-volume}
 \int_{M_8}\mathrm e^{9\lambda}\vol_8
   =2\pi^2 L^4 V_{K3}\mathcal L_z.
\end{equation}

The result agrees with the direct warped-volume integral.
The magnetic five-brane charge and the two-derivative
electric membrane charge are
\begin{align}\label{eq:k3-charge-integrals}
 N_5&=\frac{1}{(2\pi\ellp)^3}
       \frac2L(2\pi^2L^3)\mathcal L_z
       =\frac{L^2\mathcal L_z}{2\pi\ellp^3},
       \nonumber\\
 N_2^{(0)}&=\frac{1}{(2\pi\ellp)^6}
       \frac2L(2\pi^2L^3)V_{K3}
       =\frac{L^2V_{K3}}{16\pi^4\ellp^6}.
\end{align}

Consequently the leading Brown--Henneaux result is
\begin{equation}\label{eq:k3-central}
 c^{(0)}=6N_2^{(0)}N_5.
\end{equation}

There is only one Hopf generator $H$ and no trial
R-symmetry mixing. In the present normalization
$\xi=H/2$. The same magnetic transgression used in
Section~\ref{sec:physical-R-symmetry}, now with a single
Hopf weight $1/2$, reconstructs the period without
integrating the electric field:
\begin{equation}\label{eq:period-k3-charge}
 \delta=-2\pi\ell_p^3N_5.
\end{equation}

This agrees with \eqref{eq:k3-period}. The allowed
invariant closed two-form ambiguity cannot shift this
electric circle period; the cohomological argument is
given in Appendix~\ref{app:k3-curvature}.

The important difference from the first example is the
curvature of K3. Although $G_4$ has \emph{no K3 legs},
its magnetic part fills $S^3\times S^1_z$, while
$p_1(TK3)$ fills the complementary K3 directions.
Their wedge therefore produces a nonzero eight-form.
With the complex orientation of K3,
\begin{equation}\label{eq:k3-pontryagin}
 \int_{K3}p_1=-48 .
\end{equation}

Using the magnetic charge and the orientation specified
in Appendix~\ref{app:k3-curvature}, the GNS anomaly
integral becomes
\begin{equation}\label{eq:k3-anomaly-calculation}
 \int_{M_8}\mathrm e^{3\lambda}F\wedge p_1(TM_8)
 =48(2\pi\ellp)^3N_5=192\pi^2L^2\mathcal L_z .
\end{equation}

Substitution into the established $C_3\wedge X_8$
reduction \eqref{eq:anomaly} gives
\begin{equation}\label{eq:k3-anomaly-result}
 \boxed{\displaystyle
 \left.(c_L-c_R)\right|_{C_3X_8}=-12N_5.}
\end{equation}

The free-circle electric-class expression
\eqref{eq:anomaly-electric-class} reproduces the same
answer, as shown explicitly in
Appendix~\ref{app:k3-curvature}. The value is consistent
with the known higher-derivative contribution
\cite{ChowPang}; its role here is to provide a concrete
example of the sufficient obstruction test
\eqref{eq:anomaly-test}. Unlike the two-sphere product,
this magnetic obstruction is visible in the gravitational
anomaly.

The calculation determines the displayed
$C_3\wedge X_8$ contribution, not the complete quantum
values of $c_L$ and $c_R$. The superscript $(0)$ denotes
the classical two-derivative electric charge; the
shifted flux quantization and possible higher-derivative
Page corrections are discussed in
Appendix~\ref{app:k3-curvature}.

\section{\texorpdfstring{{A periodic half-BPS $\AdS_3\times S^3\times S^3$ geometry}}{A periodic half-BPS AdS3 x S3 x S3 geometry}}
\label{sec:mixed-periodic-main}

\begin{figure}[ht]
\centering
\begin{tikzpicture}[x=0.88cm,y=0.86cm,>=stealth]
 \draw[thick] (0,0)--(0,3.4)--(10,3.4)--(10,0);
 \draw[very thick] (0,3.4)--(10,3.4);
 \fill[black!25] (0,0) rectangle (3.8,0.27);
 \fill[black!4] (3.8,0) rectangle (10,0.27);
 \draw[black!70,thick] (0,0.27)--(10,0.27);
 \draw[thick,densely dashed] (5.1,0.27)--(5.1,3.4);
 \draw[->] (-0.38,0)--(-0.38,4.0) node[above] {$v$};
 \draw[->] (0,-0.7)--(10.7,-0.7) node[right] {$x$};
 \filldraw (0,0.13) circle (2.0pt);
 \filldraw (3.8,0.13) circle (2.0pt);
 \filldraw (10,0.13) circle (2.0pt);
 \node[below] at (0,-0.05) {$F_a$};
 \node[below] at (3.8,-0.05) {$F_b$};
 \node[below] at (10,-0.05) {$F_a$};
 \node at (1.85,0.74) {$q=1:\ S^3_2\to0$};
 \node at (8.1,0.74) {$q=0:\ S^3_3\to0$};
 \node[anchor=west,fill=white,inner sep=2pt]
   at (5.55,2.15) {electric cut};
 \node at (8.9,3.77) {$v=Y$};
 \node at (1.6,3.77) {outer boundary};
 \node at (3.8,-1.05) {$x=\nu T$};
 \node at (9.5,-1.05) {$x=T\sim0$};
\end{tikzpicture}
\caption{The periodic half-cylinder with
black ($q=1$) and white ($q=0$) intervals.
At $F_a,F_b$ both sphere fibers shrink and the R-symmetry
vanishes. The dashed vertical line is an arbitrary
electric-potential cut; the upper edge is the regulator
$v=Y$. The sides $x=0,T$ are identified.}
\label{fig:mixed-half-cylinder}
\end{figure}
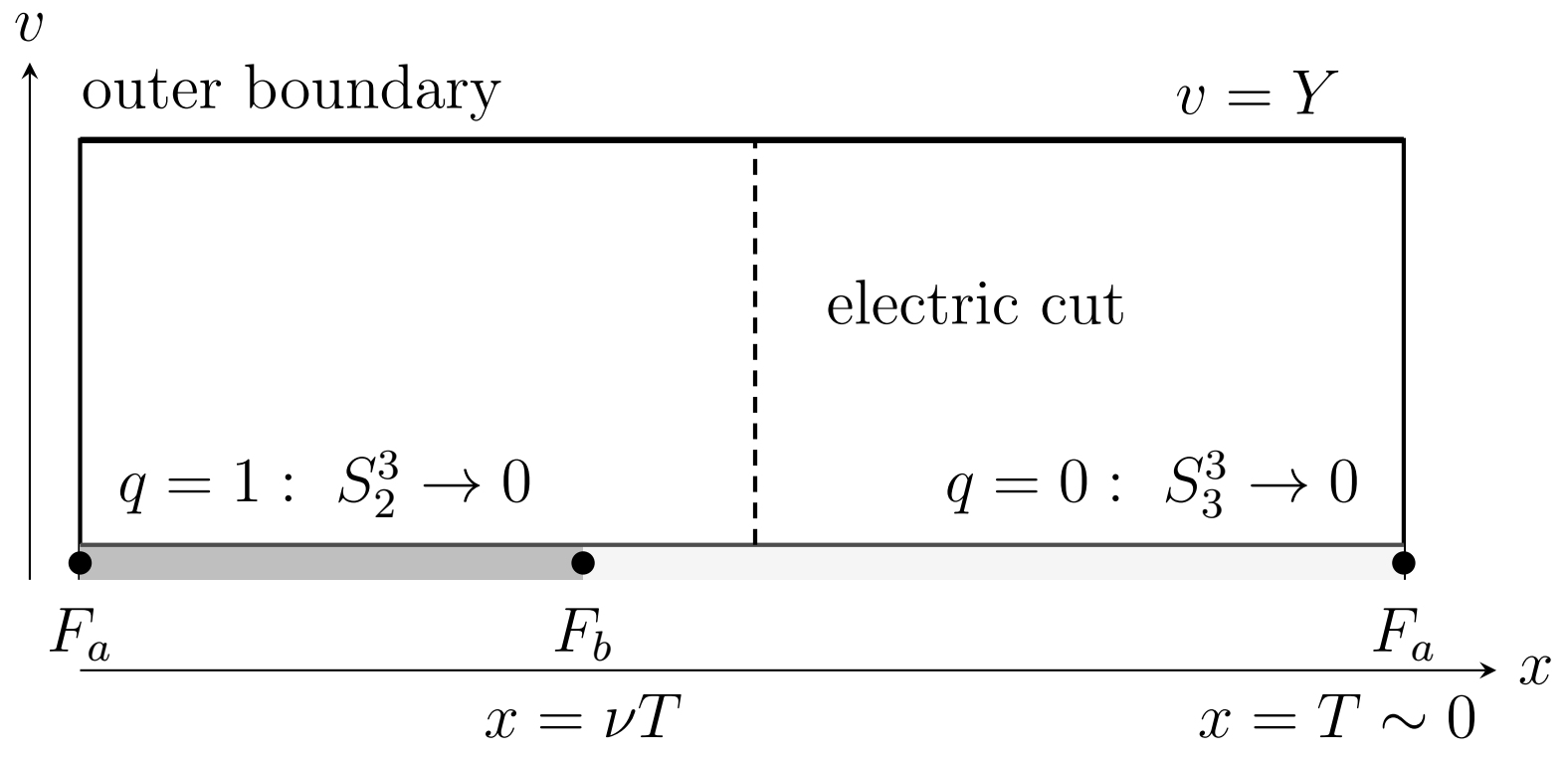

We now test the full localization formula on a smooth
half-BPS solution where all three contributions occur.
Two three-spheres are fibered over the periodic half-cylinder
$\Sigma=\{(x,v):x\sim x+T,\ v\geq0\}$.
We obtain the periodic data by periodizing the
alternating-cap configurations of
\cite[Eq.~(7.7)]{BachasDHEK2014} within the established
half-BPS family \cite{DHEGKLocal,DHEGKGlobal,ContiMacpherson2024}.
On alternating intervals of its lower edge, one or the
other sphere shrinks. At the two endpoints where the
intervals meet, both spheres shrink smoothly and the
R-symmetry has isolated fixed points. The periodic
direction carries electric monodromy, while the noncompact
end requires an outer radial regulator. We show
analytically that the three localization terms reproduce
the warped volume for \emph{every} finite cutoff.
The geometry is summarized in
Figure~\ref{fig:mixed-half-cylinder}.

\subsection{\texorpdfstring{{Geometry, fixed points and electric monodromy}}{Geometry, fixed points and electric monodromy}}

Write $w=x+\ii v$, take $0<\nu<1$, and let a fraction
$\nu$ of each period be black. The exact local BPS data
are
\begin{equation}\label{eq:mixed-bps-data}
 h=2L^3v,\qquad
 G=\frac{\ii}{2}\sum_{n\in\mathbb Z}
 \left[\frac{w-nT}{|w-nT|}
       -\frac{w-(n+\nu)T}{|w-(n+\nu)T|}\right].
\end{equation}

The terms are paired as written, so $G$ is periodic in
$x$. The BPS equation is
$\partial_wG=\tfrac12(G+\bar G)\partial_w\log h$.
Write $G=m+\ii q$ and $s=m^2+q^2$.
The eleven-dimensional metric is
\begin{equation}\label{eq:periodicmetric}
 \dd s_{11}^2=f_1^2\dd s_{\AdS_3}^2
 +f_2^2\dd s_{S^3_2}^2+f_3^2\dd s_{S^3_3}^2
 +4\rho^2(\dd x^2+\dd v^2).
\end{equation}

The warp factors are fixed algebraically by $h,G$ and
are positive for $v>0$. Their formulas, the BPS check
and the smoothness proof are given in
Appendix~\ref{app:mixed-proof}.

At $v=0$, one has $G=\ii$ on
$nT<x<(n+\nu)T$ (black intervals), where $S^3_2$
shrinks, and $G=0$ on
$(n+\nu)T<x<(n+1)T$ (white intervals), where $S^3_3$
shrinks. Both spheres shrink at the interval endpoints,
\begin{equation}\label{eq:mixed-caps}
 F_a=(0,0),\qquad F_b=(\nu T,0)
 \qquad(x\sim x+T).
\end{equation}

We call these smooth endpoints \emph{caps}. They
are not boundaries or brane singularities of $M_8$:
near either point the sphere fibrations combine into
flat $\mathbb R^4\times\mathbb R^4$,
\begin{equation}\label{eq:mixed-flat-cap}
 \dd s_{M_8}^2\simeq
  (\dd r_2^2+r_2^2\dd s_{S^3_2}^2)
 +(\dd r_3^2+r_3^2\dd s_{S^3_3}^2).
\end{equation}

The Cartesian metric, four-form and periodic Killing
spinor extend smoothly across both caps
(Appendices~\ref{app:mixed-proof}
and~\ref{app:mixed-orientation}).

Let $H_i$ be the $2\pi$-periodic Hopf generator
on the unit $S^3_i$. Raising the index of the spinor
one-form with the internal metric gives the
R-symmetry \emph{vector}
\begin{equation}\label{eq:mixed-spinor-generator-derivation}
 \xi=-\frac12H_2-H_3,
\end{equation}

whose norm is
\begin{equation}\label{eq:mixed-xi-norm}
 \|\xi\|^2=\frac{f_2^2}{4f_1^2}
                   +\frac{f_3^2}{f_1^2}.
\end{equation}

The explicit spinor bilinears are in
Appendix~\ref{app:mixed-orientation}.
Thus $\xi$ has no zeros in the interior or on the
open black/white intervals, but
$\xi(F_a)=\xi(F_b)=0$. These caps are precisely the
isolated R-symmetry fixed points needed for GNS residues.

The local electric scalar $y$ of \eqref{eq:y}
has a particularly simple differential on this solution,
\begin{equation}\label{eq:mixed-scalar-simplified}
 \dd y=2L^3(1+q)\,\dd x-2L^3m\,\dd v.
\end{equation}

This follows by exact cancellation of the
metric-dependent terms in the spinor and electric
potential expressions; the calculation is in
Appendix~\ref{app:mixed-boundary-reproduction}.
No auxiliary potential is needed here.
The mean value of $q$ over one $x$-period is $\nu$,
independent of $v$. Along the intervening black interval,
$q=1$. Thus
\begin{equation}\label{eq:mixed-electric-period-step}
 \boxed{\displaystyle
 \delta=y(x+T,v)-y(x,v)=2L^3T(1+\nu),
 \qquad y_{F_b}-y_{F_a}=4L^3\nu T.}
\end{equation}

The first quantity compares two branches at the
\emph{same} point after one circle traversal;
the second compares \emph{different} fixed points within
one branch. Neither depends on the additive origin of $y$.
The magnetic sphere-cycle periods and their quantization
are recorded in
Appendix~\ref{app:mixed-boundary-reproduction}.

\subsection{\texorpdfstring{{Fixed-point residues and the electric cut}}{Fixed-point residues and the electric cut}}

There are two distinct contributions even before
introducing the outer regulator.
First, at $F_a,F_b$ the R-symmetry vector actually
vanishes, so the standard GNS fixed-point residue applies.
Near either cap, the two shrinking three-spheres form
a locally flat $\mathbb C^2\oplus\mathbb C^2$.
Each Hopf generator acts with unit weight on both complex
coordinates of its corresponding $\mathbb C^2$.
Since $\xi=-\frac12H_2-H_3$ by
\eqref{eq:mixed-spinor-generator-derivation},
the four signed normal weights are
\begin{equation}\label{eq:mixed-signed-weights}
 \left(-\frac12,-\frac12,-1,-1\right).
\end{equation}

For the volume polyform $\Phi_0=y^3/192$.
The two caps have opposite induced orientations,
so their residues add in the physical volume
orientation to
\begin{equation}\label{eq:mixed-fixed-physical}
 I_{\rm fixed}=\frac{\pi^4}{3}
                    (y_{F_b}^3-y_{F_a}^3).
\end{equation}

The weight normalization, chirality and orientation
calculation are given in
Appendix~\ref{app:mixed-orientation}.
An additive shift of $y$ changes the individual residues;
only the complete localization sum is branch independent.

Second, the $x$-circle has electric monodromy
$y\mapsto y+\delta$. A vertical cut at any
$x_{\rm cut}$ away from the caps allows one to use a
single local branch. The resulting integral is a
jump of the same seven-form $Q_7$ across the two
sides of this \emph{auxiliary} interface, not a
fixed-point residue. By Section~\ref{sec:general},
its integrand contains terms proportional to
$\delta,\delta^2$ and $\delta^3$.

The cubic term is genuinely nonzero in this solution.
In terms of the normalized Hopf one-forms,
the R-symmetry connection is
\begin{equation}\label{eq:mixed-relative-sphere}
 \eta=-2\tau\,\eta_2-(1-\tau)\eta_3,
 \qquad
 \tau=\frac{f_2^2}{f_2^2+4f_3^2}
     =\frac{1-s}{1+s+2q}.
\end{equation}

Here $\tau$ measures the relative sizes of the two
sphere fibers. It varies radially, so
$\eta\wedge(\dd\eta)^3$ can contain one
$\dd\tau$ together with the Hopf curvature from
\emph{each} three-sphere. The cubic jump therefore
does not vanish as it did in the compact product
examples. For a cut in a white interval,
$\tau$ runs from $1$ at $v=0$ to
$(1-\nu)/(1+\nu)$ at large $v$.
The explicit Hopf contact integration is in
Appendix~\ref{app:mixed-sign-details}.

To display the two effects separately, choose the
electric branch $y_{F_a}=0$; then
$y_{F_b}=4L^3\nu T$. The fixed-point result and
the cubic part of the electric-cut contribution are
\begin{equation}\label{eq:mixed-summary}
 \boxed{\begin{aligned}
 I_{\rm fixed}
 &=\frac{64\pi^4}{3}L^9T^3\nu^3,\\
 I_{\delta^3}^{\rm white}(\infty)
 &=-\frac{32\pi^4}{3}L^9T^3\nu^2(3-\nu).
 \end{aligned}}
\end{equation}

Both are nonzero for $0<\nu<1$. The first is
localized at the two geometric caps, while the
second originates from going once around the
electric circle. A white cut has a free circle action;
a black cut has a $\mathbb Z_2$ stabilizer on its
lower edge, but the smooth eight-dimensional
boundary integral is unchanged
(Appendix~\ref{app:mixed-sign-details}).
The cut position and scalar branch need not be fixed,
because their dependence cancels against that of
the outer boundary in the full regulated expression.

\subsection{\texorpdfstring{{Exact localization at finite cutoff}}{Exact localization at finite cutoff}}
\label{sec:mixed-full-check}

The half-cylinder extends to $v\to\infty$. In that
limit $G\to\ii\nu$ and its $x$-circle shrinks in proper
length;
the metric becomes asymptotically independent
of $x$, admitting an asymptotic type IIA linear-dilaton
interpretation.
We therefore regulate the
noncompact volume by restricting to $0\leq v\leq Y$.
The number $Y>0$ is the radial integration cutoff,
not an additional parameter of the BPS solution.

Our task is to compare the direct bulk warped volume
with the sum of the three geometrically different
boundary contributions,
\begin{equation}\label{eq:mixed-regulated-target}
 \mathcal V(Y)=I_{\rm fixed}
 +I_{\rm monodromy}(Y)+I_{\rm outer}(Y).
\end{equation}

Here $I_{\rm fixed}$ comes from the two smooth caps,
$I_{\rm monodromy}$ from the jump of the electric
potential across a vertical cut, and $I_{\rm outer}$
from the regulator face at $v=Y$. In particular, the
cut is not a physical boundary of the internal space.
The warp factors \eqref{eq:mixed-warp-factors} give
$f_1f_2^3f_3^3\rho^2=L^9v(q-q^2-m^2)$.
Integrating over both unit three-spheres then gives
the bulk warped volume
\begin{equation}\label{eq:mixed-bulk-density}
 \mathcal V(Y)=16\pi^4L^9
 \int_0^T\dd x\int_0^Y\dd v\,
 v\bigl(q-q^2-m^2\bigr).
\end{equation}

Its independent Fourier--Bessel evaluation is given
in Appendix~\ref{app:mixed-analytic-endpoints}.
We now evaluate the right-hand side from the same
local seven-form $Q_7(y)$ of Section~\ref{sec:general},
using the explicit BPS fields.

The electric cut may be placed in either a white or
black interval and may be moved without changing the
answer. The jump polynomial from Section~\ref{sec:general}
splits its integral into
\begin{equation}\label{eq:mixed-cut-decomposition}
 I_{\rm monodromy}(Y)=I_\delta(Y)
              +I_{\delta^2}(Y)+I_{\delta^3}(Y).
\end{equation}

The component formulas for these three integrals and
the outer term, including their common orientation,
are in Appendix~\ref{app:mixed-boundary-reproduction}.
They allow an analytic check that goes beyond a
particular numerical cutoff.

Indeed, the BPS equations
$m_v=q_x+m/v$ and $q_v=-m_x$, together with
the spinorial scalar identity
\eqref{eq:mixed-scalar-simplified}, imply an explicit
differential identity for the sphere-integrated
seven-form. As a result, the \emph{derivative} of the
complete boundary sum is
\begin{align}\label{eq:mixed-boundary-derivative}
 \frac{\dd}{\dd Y}\left[
 I_{\rm fixed}+I_{\rm monodromy}(Y)+I_{\rm outer}(Y)
 \right]
 &=16\pi^4L^9Y\int_0^T
       \bigl(q-q^2-m^2\bigr)\,\dd x\nonumber\\
 &=\frac{\dd\mathcal V}{\dd Y}.
\end{align}

The equality on the first line is obtained by
differentiating the \emph{explicit} spinor-bilinear
coefficients of $Q_7$, with the $y$, $y^2$ and $y^3$
terms cancelling separately
(Appendix~\ref{app:mixed-analytic-endpoints}).
It is not obtained by substituting the known bulk
volume into the boundary result.

Equality of derivatives fixes the answer up to a
constant. That constant is determined independently
at the smooth caps. As $Y\to0^+$, the electric-cut
integral vanishes, while the regulated outer face
collapses onto the two cap contributions with
the opposite orientation,
\begin{equation}\label{eq:mixed-bottom-boundary}
 \lim_{Y\to0^+}I_{\rm monodromy}(Y)=0,\qquad
 \lim_{Y\to0^+}I_{\rm outer}(Y)=-I_{\rm fixed}.
\end{equation}

Since $\mathcal V(0)=0$, the integration constant
vanishes. We thus establish, for arbitrary allowed
cut position and electric scalar branch,
\begin{equation}\label{eq:mixed-all-cutoffs}
 \boxed{\displaystyle
 \mathcal V(Y)=I_{\rm fixed}+I_\delta(Y)
  +I_{\delta^2}(Y)+I_{\delta^3}(Y)
  +I_{\rm outer}(Y),\qquad Y>0.}
\end{equation}

This is an identity at \emph{every} finite radial
cutoff, not a numerical agreement at a chosen value
of $Y$. The exact cut primitives, the lower-cap limit
and the cancellation of cut and branch dependence
are derived in
Appendix~\ref{app:mixed-analytic-endpoints}.

Finally, the large-$Y$ behavior has a simple physical
interpretation. The direct bulk integration gives
\begin{equation}\label{eq:mixed-bulk-large-Y}
 \begin{aligned}
 \mathcal V(Y)
 ={}&8\pi^4L^9T\nu(1-\nu)Y^2\\
 &-\frac43\pi^4L^9T^3\nu^2(1-\nu)^2
 +O\!\left(Y^2e^{-4\pi Y/T}\right).
 \end{aligned}
\end{equation}

The boundary computation \emph{independently} yields
$I_\delta(Y)=8\pi^4L^9T\nu(1-\nu)Y^2+O(1)$;
the quadratic and cubic cut terms, fixed-point
residues and outer boundary are all $O(1)$
(see \eqref{eq:mixed-outer-limit} for the latter).
Thus the entire quadratic divergence is carried by
the term \emph{linear} in the electric period.
The finite pieces also combine to the constant in
\eqref{eq:mixed-bulk-large-Y}, with cut-location
and branch dependence cancelling analytically
(Appendix~\ref{app:mixed-analytic-endpoints}).
The outer contribution is generically nonzero in
the large-$Y$ limit, although its leading finite
value vanishes at the special parameter $\nu=1/3$.

Because the noncompact end has linear-dilaton
asymptotics, this result is a statement about the
\emph{regulated} warped volume. A finite holographic
central charge at that end would require a separate
boundary renormalization prescription.

\section{Discussion}\label{sec:discussion}

{%
The problem addressed here is global. The GNS volume form
\cite{GNS} has a local primitive away from the R-symmetry
fixed set, and this primitive depends on an electric scalar
that is single valued only when the electric one-form is
exact. Stokes' theorem on the electric-potential patches
gives \eqref{eq:modified-gns-formula}, in which fixed-point
residues, electric jumps and outer boundaries have distinct
geometric origins. The jumps admit the closed quotient
four-form $g_\delta$ of \eqref{eq:gdelta-pullback}, so a cut
contribution on a free R-symmetry quotient is a
cohomological pairing, and the full sum is independent of
the admissible cuts and of the electric branch. Smooth
compact solutions with a nowhere-vanishing R-symmetry vector
and positive warped volume therefore carry non-exact
electric flux.

The physical statements of Section~\ref{sec:physical} rest
on additional hypotheses, which we kept separate. For a
compact free R-action with a single electric period, the
integrated quadratic and cubic cut terms cancel, and on the
$SU(3)$ spinor branch the surviving term is the electric
Maxwell charge, related to the Page charge by
\eqref{eq:rank-one-page}. Independently, supersymmetry links
the electric one-form to the circle pushforward of the
magnetic four-form, \eqref{eq:gysin-pushforward}. The
product of the Euler class with the electric class decides
whether the magnetic flux has a global ordinary equivariant
completion, and its pairing with $p_1$ is the
$C_3\wedge X_8$ contribution \eqref{eq:anomaly-electric-class}.
A nonzero anomaly contribution therefore detects the
obstruction, while the localization formula
\eqref{eq:modified-gns-formula} holds in all cases.

The compact examples test these statements in complementary
ways. The $S^3\times S^3\times T^2$ solution has a nonzero
electric period and magnetic obstruction but vanishing
$p_1$, so its $C_3\wedge X_8$ contribution vanishes.
Magnetic transgression and a Page-fixed eleven-dimensional
electric-action calculation give the same polynomial in the
trial R-symmetry, the Killing spinor selects its stationary
point, and the resulting level and central charge agree with
the sphere-isometry results \cite{Gukov2004,HST2025}. The
current level is nonzero although $G_4\wedge G_4=0$, because
it comes from the Page-fixed kinetic term. The
$S^3\times K3\times S^1$ solution has a nonzero Pontryagin
class and detects its magnetic obstruction through the
anomaly. Reference \cite{GregoryAnomaly} localizes the
$C_3\wedge X_8$ trial anomaly on projective bundles over a
six-dimensional base, for magnetically flux-quantized
backgrounds, and confirms the mixed coefficient by direct
fibre integration. Equation \eqref{eq:anomaly-electric-class}
covers free R-symmetry actions, which have no fixed points,
and expresses the same contribution through the electric
class. Combining the two descriptions on backgrounds that
have both fixed points and non-exact electric flux is a
natural next step.

The periodic half-BPS example shows that the fixed-point,
electric-cut and outer-boundary terms can all be nonzero in
one smooth supergravity solution. Its regulated bulk volume
has the closed form \eqref{eq:mixed-analytic-bulk-volume}.
This example is a periodic realization within
the known large $\mathcal N=(4,4)$ local family. A
classification of globally flux-quantized periodic
quotients is beyond the scope of this work.
The explicit BPS boundary forms obey a coordinate differential
identity that proves equality with the fixed-point, cut and
outer-boundary sum at every finite radial cutoff; the cap limit
fixes its integration constant independently. The individual
cut orders also have analytically determined large-cutoff
behaviour, with the full quadratic divergence arising from
the term linear in the electric period. The linear-dilaton
end of the geometry requires
a separate boundary prescription before a finite holographic
central charge can be assigned.

One direction for further study is to find a
smooth compact supersymmetric background in which both the
R-symmetry fixed-point set and nontrivial electric periods
are present. In such a case the fixed-point residues and
electric-cut contributions would coexist without an outer
boundary, providing a direct global test independent of
noncompact boundary conditions. The patch formula already
allows several independent electric periods and relative
magnetic Page data, but the Maxwell-charge reduction used
here assumes a single period and a particular $SU(3)$
spinor structure. Determining how distinct periods pair
with Maxwell and Page charges beyond this rank-one setting
is a separate problem.

For noncompact backgrounds, an important next
step is to establish a holographically appropriate
renormalization prescription for the warped volume and to
determine whether its finite part can be expressed through
renormalized fixed-point, cut and outer-boundary data.
The periodic example establishes the regulated identity
but does not itself assign a finite central charge to the
linear-dilaton end. Higher-derivative corrections to Page
charges and to the separate central charges $c_L$ and
$c_R$, as well as the relation to the full M-theory path
integral \cite{MTheoryPath,WittenFlux}, remain further
directions for study.
}

\section*{Acknowledgements}
The author acknowledges the assistance of OpenAI's GPT-6 Astra
in developing and checking the analytic proof of the finite-cutoff
identity in Section~\ref{sec:mixed-full-check}. It also assisted
with literature searches, parts of the intermediate
calculations, and manuscript drafting and editing. The research
ideas, physical concepts, overall direction and final scientific
decisions were determined by the author. All final derivations,
calculations, statements and conclusions were independently
examined and verified by the author, who takes full responsibility
for the content of this work.

\appendix

\section{Global descent and electric monodromy}\label{app:patches}

\subsection{Local equivariant closure}
\label{app:physical-pairing-proof}

{%
We first display the componentwise descents of $\Phi^\Psi$
and $\Phi$ quoted in Section~\ref{sec:review}. Each line
uses \eqref{eq:bilinear}, \eqref{eq:contractions} and
$\dd y=\mathrm e^{3\lambda}K$.
}
Since $\mathrm e^{6\lambda}\Psi$ is not closed,
its four-form completion requires $y\mathrm e^{3\lambda}F${,}
\begin{align}
 \dd\left(\mathrm e^{6\lambda}\Psi
              +y\mathrm e^{3\lambda}F\right)
 &=-\mathrm e^{3\lambda}K\wedge\mathrm e^{3\lambda}F
  +\mathrm e^{3\lambda}K\wedge\mathrm e^{3\lambda}F=0,
 \nonumber\\
 \dd\left(-\frac y2\mathrm e^{3\lambda}\omega\right)
 &=-\frac12\mathrm e^{3\lambda}K\wedge
                       \mathrm e^{3\lambda}\omega
   +y\iota_\xi(\mathrm e^{3\lambda}F)
  \nonumber\\
 &=\iota_\xi\left(\mathrm e^{6\lambda}\Psi
                +y\mathrm e^{3\lambda}F\right),
 \nonumber\\
 \dd\left(-\frac{y^2}{8}\right)
 &=-\frac y4\mathrm e^{3\lambda}K
  =\iota_\xi\left(-\frac y2\mathrm e^{3\lambda}\omega\right).
\end{align}

For the warped volume, the remaining descent is
\begin{align}
 \dd\left(\frac14\mathrm e^{9\lambda}*J\right)
 &=\iota_\xi\left(\mathrm e^{9\lambda}\vol_8\right),
 \nonumber\\
 \dd\left(-\frac y8\mathrm e^{6\lambda}\Psi
              -\frac{y^2}{16}\mathrm e^{3\lambda}F\right)
 &=-\frac18\mathrm e^{3\lambda}K\wedge
                       \mathrm e^{6\lambda}\Psi
   +\left(\frac y8-\frac y8\right)
      \mathrm e^{3\lambda}K\wedge\mathrm e^{3\lambda}F
   \nonumber\\
 &=-\frac18\mathrm e^{3\lambda}K\wedge
                    \mathrm e^{6\lambda}\Psi
   =\iota_\xi\left(\frac14\mathrm e^{9\lambda}*J\right),
 \nonumber\\
 \dd\left(\frac{y^2}{32}\mathrm e^{3\lambda}\omega\right)
 &=\frac y{16}\mathrm e^{3\lambda}K\wedge
                         \mathrm e^{3\lambda}\omega
   -\frac{y^2}{16}\iota_\xi(\mathrm e^{3\lambda}F)
   \nonumber\\
 &=\iota_\xi\left(-\frac y8\mathrm e^{6\lambda}\Psi
             -\frac{y^2}{16}\mathrm e^{3\lambda}F\right),
 \nonumber\\
 \dd\left(\frac{y^3}{192}\right)
 &=\frac{y^2}{64}\mathrm e^{3\lambda}K
   =\iota_\xi\left(\frac{y^2}{32}
                       \mathrm e^{3\lambda}\omega\right).
\end{align}
The displayed cancellations establish the local
equivariant closure \eqref{eq:local-closure} directly from
the supergravity bilinear relations. For completeness,
the remaining magnetic completion obeys
\begin{equation}
 \dd\left(-\frac12\mathrm e^{3\lambda}\omega\right)
 =\iota_\xi(\mathrm e^{3\lambda}F),\qquad
 \dd\left(-\frac y4\right)
 =\iota_\xi\left(-\frac12\mathrm e^{3\lambda}\omega\right).
\end{equation}
All three local forms consequently transform by
\eqref{eq:monodromy} when the electric scalar changes
branch. Their local closure and ordinary Stokes'
theorem suffice for the global localization formula.
\subsection{The local primitive and closed-cut pairing}
\label{app:cut-pairing}

On the locus where $\xi\neq0$, let $\eta$ and
$\Omega=\dd\eta$ be as in \eqref{eq:eta}. Since
$\iota_\xi\eta=1$, $\iota_\xi\Omega=0$ and $\Omega$
is a two-form on the seven-dimensional horizontal
distribution, $\Omega^4=0$. The finite equivariant
homotopy is therefore
\begin{equation}\label{eq:appendix-direct-homotopy}
 h_\xi=-\eta\wedge(1+\Omega+\Omega^2+\Omega^3),
 \qquad
 \dd_\xi h_\xi
 =(\Omega-1)\bigl[-(1+\Omega+\Omega^2+\Omega^3)\bigr]
 =1.
\end{equation}
For the closed GNS polyform $\dd_\xi\Phi(y)=0$,
taking the ordinary eight-form component of
$\dd_\xi(h_\xi\wedge\Phi)=\Phi$ yields
\begin{equation}
 Q_7(y)=[h_\xi\wedge\Phi(y)]_7,\qquad
 \dd Q_7(y)=\mathrm e^{9\lambda}\vol_8.
\end{equation}
This is exactly \eqref{eq:Q7}--\eqref{eq:dQ}.
At a fixed component $\mathcal F$, $\eta$ is undefined;
applying Stokes' theorem after excising a small normal
disc bundle gives \eqref{eq:fixed-residue} in the
shrinking-tube limit.

We now verify the quotient statement for a closed cut
$\mathcal C$ with a free R-symmetry circle of period
$\ell$. Write $\pi:\mathcal C\to\mathcal B_6$ and
$\Omega=\pi^*\bar\Omega$. For an invariant differential
form $\alpha$, its horizontal projection
$\alpha_h=\alpha-\eta\wedge\iota_\xi\alpha$ is pulled
back from the quotient. The projections entering
\eqref{eq:gdelta-pullback} are
\begin{align}\label{eq:gdelta-horizontal-projections}
 \bigl(\mathrm e^{6\lambda}\Psi\bigr)_h
 &=\mathrm e^{6\lambda}\Psi
  +\tfrac12\eta\wedge\mathrm e^{3\lambda}K
                \wedge\mathrm e^{3\lambda}\omega,
 \nonumber\\
 \bigl(\mathrm e^{3\lambda}F
    -\tfrac12\Omega\wedge\mathrm e^{3\lambda}\omega\bigr)_h
 &=\mathrm e^{3\lambda}F
  -\eta\wedge\iota_\xi(\mathrm e^{3\lambda}F)
  -\tfrac12\Omega\wedge\mathrm e^{3\lambda}\omega
  +\tfrac14\eta\wedge\Omega\wedge\mathrm e^{3\lambda}K.
\end{align}
The spinor identities \eqref{eq:bilinear} and
\eqref{eq:contractions}, together with
$\dd\Omega=0$, give
\begin{align}\label{eq:gdelta-differentials}
 \dd\bigl(\mathrm e^{6\lambda}\Psi\bigr)_h
 &=-\mathrm e^{3\lambda}K\wedge
  \bigl(\mathrm e^{3\lambda}F
    -\tfrac12\Omega\wedge\mathrm e^{3\lambda}\omega\bigr)_h,
 \nonumber\\
 \dd\bigl(\mathrm e^{3\lambda}F
    -\tfrac12\Omega\wedge\mathrm e^{3\lambda}\omega\bigr)_h
 &=\tfrac14\Omega^2\wedge\mathrm e^{3\lambda}K.
\end{align}
Differentiating \eqref{eq:gdelta-pullback} and using
$\dd y=\mathrm e^{3\lambda}K$ now gives the explicit
cancellation
\begin{align}\label{eq:gdelta-closed}
 \dd(\pi^*g_\delta)={}&
 \left(\frac{\delta}{8}-\frac{2\delta}{16}\right)
 \mathrm e^{3\lambda}K\wedge
 \bigl(\mathrm e^{3\lambda}F
    -\tfrac12\Omega\wedge\mathrm e^{3\lambda}\omega\bigr)_h
 \nonumber\\
 &+\left[-\frac{2y\delta+\delta^2}{64}
      +\frac{6y\delta+3\delta^2}{192}\right]
 \mathrm e^{3\lambda}K\wedge\Omega^2=0.
\end{align}
Thus $\dd g_\delta=0$. Substitution into the
jump of $Q_7$ and integration over the R-symmetry
circle give
\begin{equation}\label{eq:app-cut-cohomological-pairing}
 \int_{\mathcal C}\Delta_\delta Q_7
 =-\ell\int_{\mathcal B_6}\bar\Omega\wedge g_\delta
 =-\ell^2\int_{\mathcal B_6}e\wedge g_\delta,
 \qquad e=[\bar\Omega/\ell],
\end{equation}
as in \eqref{eq:cut-cohomological-pairing}.
For a cut with edges, its contribution must instead
be combined with the outer-boundary terms of
\eqref{eq:modified-gns-formula}.
\section{\texorpdfstring{{Spinorial identities for the charge and anomaly relations}}{Spinorial identities for the charge and anomaly relations}}
\label{app:spinorial-identities}

\subsection{The SU(3) spinorial pairing}
\label{app:maxwell-spinor-proof}

We give the local Killing-spinor calculation underlying
\eqref{eq:maxwell-bilinear-pairing}. On the $SU(3)$ branch
with vanishing scalar bilinears $\zeta,S,P$ in the
classification of \cite{Ashmore}, choose an oriented
orthonormal frame such that
\begin{equation}\label{eq:su3-normalform}
 K=-e^7,\qquad \xi^\flat=\frac12e^3,\qquad
 J=e^{12}+e^{45}+e^{68},\qquad
 \omega=-e^{37},\qquad \Psi=-\frac12J^2.
\end{equation}
The one-form projection of the algebraic supersymmetry
equation is
\begin{equation}\label{eq:su3-kse-projection}
 J\mathbin{\lrcorner}\iota_\xi F+f+3K=0.
\end{equation}
To see its coefficient, write
$\not\!F=F_{abcd}\gamma^{abcd}/4!$ and use
the algebraic Killing-spinor operator
\begin{equation}
 \mathcal M=\frac12\partial_a\lambda\,\gamma^a
 -\frac1{12}\not\!F
 -\frac16f_a\gamma^a\gamma_9+\frac12\gamma_9,
 \qquad\mathcal M\chi=0.
\end{equation}
In the frame \eqref{eq:su3-normalform}, the Fierz
contractions give
\begin{equation}
 \operatorname{Re}(\chi^\dagger\gamma_a\gamma_9
                    \gamma_b\chi)=0,\qquad
 \operatorname{Re}(\chi^\dagger\gamma_a\gamma_9
                     \not\!F\chi)
 =-2(J\mathbin{\lrcorner}\iota_\xi F)_a.
\end{equation}
The remaining contractions needed here are
\begin{equation}
 \operatorname{Re}(\chi^\dagger\gamma_a\gamma^b\chi)
 =\delta_a{}^b,\qquad
 \operatorname{Re}(\chi^\dagger\gamma_a\chi)=K_a,
 \qquad \gamma_9\gamma^b\gamma_9=-\gamma^b.
\end{equation}
Multiplying $\mathcal M\chi=0$ by
$6\operatorname{Re}(\chi^\dagger\gamma_a\gamma_9\,\cdot\,)$
then gives the four contributions in their KSE order,
\begin{equation}\label{eq:su3-kse-four-terms}
 0=\underbrace{0}_{\dd\lambda}
 +\underbrace{(J\mathbin{\lrcorner}\iota_\xi F)_a}_{F}
 +\underbrace{f_a}_{f}
 +\underbrace{3K_a}_{\gamma_9}.
\end{equation}
This proves \eqref{eq:su3-kse-projection} with unrestricted
magnetic and electric flux components.

{For the differential step note that $2\xi^\flat=e^3$ and,
since $\|\xi\|^2=1/4$, $\eta=2e^3$.}
The differential Killing-spinor identity in the present
normalization is
\begin{equation}\label{eq:su3-differential-kse}
 \mathrm e^{-3\lambda}\dd(\mathrm e^{3\lambda}{e^3})
 =J-\frac12\omega\mathbin{\lrcorner} F
   +\frac12J\mathbin{\lrcorner}*F.
\end{equation}
The term involving $\dd\lambda\wedge{e^3}$
drops from ${e^3}\wedge\dd{e^3}$.
Consequently,
\begin{align}\label{eq:su3-wedge-expansion}
 \eta\wedge\dd\eta\wedge\Psi
 &=4{e^3}\wedge\dd{e^3}\wedge\Psi\nonumber\\
 &=4{e^3}\wedge
  \left(J-\frac12\omega\mathbin{\lrcorner} F
       +\frac12J\mathbin{\lrcorner}*F\right)
       \wedge\Psi.
\end{align}
Using $\Psi=-J^2/2$ and the frame
\eqref{eq:su3-normalform}, the flux-independent and
flux-dependent terms obey, respectively,
\begin{align}
 4{e^3}\wedge J\wedge\Psi
 &=-12*_{\rm GNS}K,\nonumber\\
 2{e^3}\wedge
   \left(J\mathbin{\lrcorner}*F
        -\omega\mathbin{\lrcorner}F\right)\wedge\Psi
 &=-4*_{\rm GNS}
      \left(J\mathbin{\lrcorner}\iota_\xi F\right).
\end{align}
These are wedge--Hodge identities in the displayed frame;
they hold componentwise for a general four-form $F$.
Adding the two terms yields
\begin{equation}\label{eq:su3-pre-maxwell}
 \eta\wedge\dd\eta\wedge\Psi
 =-4*_{\rm GNS}
       \left(J\mathbin{\lrcorner}\iota_\xi F+3K\right).
\end{equation}
Finally \eqref{eq:su3-kse-projection} reduces this
seven-form to $4*_{\rm GNS}f$. Multiplication by
$\mathrm e^{6\lambda}$ proves
\eqref{eq:maxwell-bilinear-pairing}.
The calculation uses the normal form of the spinor,
with unrestricted magnetic and electric flux components.
The sign follows from the GNS convention
$*_{\rm GNS}e^7=-e^{1234568}$.

\subsection{Magnetic completion and orientations}
\label{app:magnetic-completion}

The representative freedom in
\eqref{eq:magnetic-electric-class} follows from the
magnetic Killing-spinor descent. For an invariant
three-form $\Lambda$ and an invariant closed two-form
$\vartheta_2$, consider the formal representative changes
\begin{align}\label{eq:magnetic-representative-shift}
 (\mathrm e^{3\lambda}F)'
 &=\mathrm e^{3\lambda}F+\dd\Lambda,\nonumber\\
 (\mathrm e^{3\lambda}\omega)'
 &=\mathrm e^{3\lambda}\omega
   +2\iota_\xi\Lambda+\vartheta_2,
 \qquad \dd\vartheta_2=0,
 \qquad \mathcal L_\xi\Lambda
 =\mathcal L_\xi\vartheta_2=0.
\end{align}
Cartan's identity gives
$\dd(\iota_\xi\Lambda)=-\iota_\xi\dd\Lambda$;
thus
\begin{equation}
 \dd(\mathrm e^{3\lambda}\omega)'
 =-2\iota_\xi(\mathrm e^{3\lambda}F+\dd\Lambda)
 =-2\iota_\xi(\mathrm e^{3\lambda}F)'.
\end{equation}
Using $\iota_\xi^2=0$ and
$\mathrm e^{3\lambda}K=2\iota_\xi(\mathrm e^{3\lambda}\omega)$
then gives
\begin{equation}\label{eq:period-ambiguity}
 (\mathrm e^{3\lambda}K)'-\mathrm e^{3\lambda}K
 =2\iota_\xi\vartheta_2,
 \qquad
 [(\mathrm e^{3\lambda}f)']-[\mathrm e^{3\lambda}f]
 =-4[\iota_\xi\vartheta_2].
\end{equation}
The last class is well defined because
$\dd(\iota_\xi\vartheta_2)
=\mathcal L_\xi\vartheta_2-\iota_\xi\dd\vartheta_2=0$.
{These shifts compare solutions of the descendant
equations with the same magnetic cohomology class within one
supersymmetric background, in which the Killing spinor fixes
$\omega$.}

{%
For a free $\xi$ action, the Gysin sequence reads
\begin{equation}
 H^4(B_7)\xrightarrow{\pi^*}H^4(M_8)
 \xrightarrow{\pi_*}H^3(B_7)
 \xrightarrow{e\wedge}H^5(B_7).
\end{equation}
The equivariant cohomology of a free action is the ordinary
cohomology of $B_7$, so a closed magnetic four-form has an
ordinary global Cartan completion precisely when its class
lies in the image of $\pi^*$, that is when
$\pi_*[\mathrm e^{3\lambda}F]=0$.

For completeness we also derive the obstruction directly
from the magnetic descent. Keeping the top component
fixed and allowing an invariant closed two-form
$\vartheta_2$, consider

\begin{equation}\label{eq:other-completion}
 \mathrm e^{3\lambda}F
 -\frac12\left(\mathrm e^{3\lambda}\omega+\vartheta_2\right)
 -\frac14\,y',
 \qquad
 \dd y'=\mathrm e^{3\lambda}K+2\iota_\xi\vartheta_2,
 \qquad \dd\vartheta_2=0.
\end{equation}

This polyform is $\dd_\xi$-closed by
\eqref{eq:magnetic-descent}; a global scalar descendant
$y'$ exists exactly when the shifted one-form on the
right is exact. An invariant closed two-form decomposes
along the circle bundle as

\begin{equation}\label{eq:vartheta-decomposition}
 \vartheta_2=\pi^*\bar\vartheta_2
 +\eta\wedge\pi^*\bar\vartheta_1,
 \qquad
 \dd\bar\vartheta_1=0,\qquad
 \dd\bar\vartheta_2=-\bar\Omega\wedge\bar\vartheta_1.
\end{equation}

Hence $e\wedge[\bar\vartheta_1]=0$ and
$\iota_\xi\vartheta_2=\pi^*\bar\vartheta_1$.
An invariant primitive of the horizontal one-form
$\mathrm e^{3\lambda}K+2\iota_\xi\vartheta_2$
descends to $B_7$. Consequently a global $y'$ requires
$[\overline{\mathrm e^{3\lambda}K}]
=-2[\bar\vartheta_1]$, so
$e\wedge[\overline{\mathrm e^{3\lambda}K}]=0$.
Conversely, when the latter class vanishes, one may
choose $\bar\vartheta_1
=-\overline{\mathrm e^{3\lambda}K}/2$ and solve
$\dd\bar\vartheta_2=-\bar\Omega\wedge\bar\vartheta_1$.
This gives a globally defined completion, proving the
Gysin criterion directly.

The coefficient of this obstruction follows from the
spinor bilinears. The horizontal part of the two-form is
$(\mathrm e^{3\lambda}\omega)_h
=\mathrm e^{3\lambda}\omega
-\frac12\eta\wedge\mathrm e^{3\lambda}K$.
Differentiating with $\dd(\mathrm e^{3\lambda}K)=0$
gives

\begin{equation}\label{eq:gysin-local-derivation}
 \iota_\xi(\mathrm e^{3\lambda}F)
 =-\frac12\dd(\mathrm e^{3\lambda}\omega)_h
  -\frac14\Omega\wedge\mathrm e^{3\lambda}K,
 \qquad
 \pi_*(\mathrm e^{3\lambda}F)
 =\ell_\xi\,\overline{\iota_\xi(\mathrm e^{3\lambda}F)}.
\end{equation}

The first term is exact on $B_7$; using
$[\bar\Omega]=\ell_\xi e$ and
$[\mathrm e^{3\lambda}f]
=-2[\mathrm e^{3\lambda}K]$ proves
\eqref{eq:gysin-pushforward}. Since the R-symmetry
circle contributes a trivial real line to the tangent
bundle, $p_1(TM_8)=\pi^*p_1(TB_7)$, so fibre integration
gives

\begin{equation}\label{eq:anomaly-pushforward}
 \int_{M_8}\mathrm e^{3\lambda}F\wedge p_1(TM_8)
 =\int_{B_7}\pi_*(\mathrm e^{3\lambda}F)\wedge p_1(TB_7).
\end{equation}

Substituting \eqref{eq:gysin-pushforward} into
\eqref{eq:anomaly-pushforward} gives
\eqref{eq:anomaly-electric-class}, a rewriting of the
GNS anomaly formula \eqref{eq:anomaly}.

In cohomological language the magnetic second descendant
determines $[\mathrm e^{3\lambda}f]/8$ in the quotient of
$H^1(M_8;\mathbb R)$ by the classes $[\iota_\xi\vartheta_2]$ with
$\dd\vartheta_2=0$ and $\mathcal L_\xi\vartheta_2=0$, and the direct
period statement is \eqref{eq:magnetic-electric-class}.

}

\paragraph{Hopf reconstruction of the electric period.}
For the $S^3_+\times S^3_-\times T^2$ background, the
magnetic descent without its local scalar component is

\begin{equation}\label{eq:period-spinor-descent}
 \dd_\xi\Bigl(\mathrm e^{3\lambda}F
 -\frac12\mathrm e^{3\lambda}\omega\Bigr)
 =\frac14\mathrm e^{3\lambda}K,\qquad
 \mathcal L_\xi\chi=\frac{\mathrm i}{2}\chi .
\end{equation}

Normalize the two Hopf generators $H_\pm$ to have period
$2\pi$, and choose one-forms $\eta_\pm$ with
$\eta_\pm(H_\pm)=1$ and
$\int_{S^3_\pm}\eta_\pm\wedge\dd\eta_\pm=4\pi^2$.
The corresponding classes are
$a_\pm=\eta_\pm\wedge\dd\eta_\pm/(4\pi^2)$.
Writing $s_z=\dd z/\mathcal L_z$, with unit integral,
the magnetic class is
$[\mathrm e^{3\lambda}F]=(2\pi\ell_p)^3
\left(N_5^+[a_+\wedge s_z]+N_5^-[a_-\wedge s_z]\right)$.
For $\xi=w_+H_++w_-H_-$, Cartan's formula and
$\iota_{H_\pm}\dd\eta_\pm=0$ give

\begin{equation}\label{eq:hopf-transgression}
 \dd_\xi\Bigl(a_\pm+\frac{w_\pm}{4\pi^2}\eta_\pm\Bigr)
 =-\frac{w_\pm^2}{4\pi^2}.
\end{equation}

The two-form contributions cancel between the two
terms. Multiplying the completions by the magnetic
charges and $s_z$ leaves the residual one-form

\begin{equation}\label{eq:hopf-residual-cohomology}
 \dd_\xi\Bigl[(2\pi\ell_p)^3\sum_{\pm}N_5^\pm
 \Bigl(a_\pm+\frac{w_\pm}{4\pi^2}\eta_\pm\Bigr)
 \wedge s_z\Bigr]
 =-\frac{(2\pi\ell_p)^3}{4\pi^2}
 \left(N_5^+w_+^2+N_5^-w_-^2\right)s_z .
\end{equation}

Its top component represents the same magnetic
cohomology class as $\mathrm e^{3\lambda}F$.
The descendants can therefore differ by an invariant
equivariantly exact representative and by the closed
two-form freedom $\vartheta_2$ of
\eqref{eq:magnetic-representative-shift}. The $S^1_z$
period of $\iota_\xi\vartheta_2$ vanishes by the argument
below. Comparing \eqref{eq:period-spinor-descent} with
\eqref{eq:hopf-residual-cohomology} gives

\begin{equation}\label{eq:hopf-period}
 \delta=\oint_{S^1_z}\mathrm e^{3\lambda}K
 =-\frac{(2\pi\ell_p)^3}{\pi^2}
  \left(N_5^+w_+^2+N_5^-w_-^2\right)
 =-8\pi\ell_p^3
  \left(N_5^+w_+^2+N_5^-w_-^2\right).
\end{equation}

For $\xi_t=R_t/2$ the weights are $w_+=t/2$ and
$w_-=(1-t)/2$, yielding \eqref{eq:period-trial}.
For irrational weight ratios, the orbits of $\xi_t$
may be dense; the formula is then a transgression
polynomial on the torus Lie algebra, not the period of
a distinct BPS solution. The physical period is its
value at the spinor-selected weights.

\paragraph{The $S^3_+\times S^3_-\times T^2$ obstruction.}
\label{app:magnetic-completion-example}
{%
The completion \eqref{eq:other-completion} allows an
arbitrary invariant closed two-form $\vartheta_2$. Since
\begin{equation}
 H^2(S^3_+\times S^3_-\times T^2;\mathbb R)
 \cong H^2(T^2;\mathbb R),
\end{equation}
every such $\vartheta_2$ differs from a constant multiple of
$\dd z\wedge\dd w$ by an exact form $\dd\Lambda$, and the
one-form $\Lambda$ can be chosen invariant by averaging over
the compact isometry group. The Killing vector $\xi$ is
tangent to the two three-spheres, so
$\iota_\xi(\dd z\wedge\dd w)=0$, and Cartan's identity gives
\begin{equation}
 \iota_\xi\vartheta_2=\iota_\xi\dd\Lambda
 =-\dd(\iota_\xi\Lambda),
 \qquad
 \oint_{S^1_z}\iota_\xi\vartheta_2=0.
\end{equation}
On the other hand \eqref{eq:spheres-period} gives
$\oint_{S^1_z}\mathrm e^{3\lambda}K=\delta\ne0$. The $S^1_z$
period of $\dd y'=\mathrm e^{3\lambda}K+2\iota_\xi\vartheta_2$ is
therefore $\delta$ for every $\vartheta_2$, and the magnetic flux
has no globally defined ordinary completion. The localization
formula of Section~\ref{sec:general} holds regardless. The
product tangent bundle has $p_1(TM_8)=0$, so the anomaly
pairing \eqref{eq:anomaly} vanishes although the completion
obstruction is nonzero.
}
\section{Electric polarization and bulk current levels}
\label{app:bulk-current}

For the $S^3_+\times S^3_-\times T^2$ background,
write $s_z=\dd z/\mathcal L_z$. Both the electric
and magnetic four-form components contain this factor,
so $G_4\wedge G_4=0$. The parity-odd three-dimensional
current coupling instead follows by gauging the sphere
isometries in the electric kinetic term and fixing its
Page momentum. We give the steps underlying the
independent check in Section~{\ref{sec:physical-R-symmetry}}.

Each sphere has $SO(4)\simeq(SU(2)_L\times SU(2)_R)/\mathbb Z_2$
isometry, with Kaluza--Klein connections
$\mathcal A_{\pm,L}$ and $\mathcal A_{\pm,R}$.

Let $e_{3,\pm}$ be normalized angular forms on the two
three-spheres, and let $\Theta_{3,\pm}$ be their Euler
Chern--Simons descendants,
\begin{align}\label{eq:bulk-angular-forms}
 &\int_{S^3_\pm}e_{3,\pm}=1,\qquad
 \dd e_{3,\pm}=\dd\Theta_{3,\pm},\nonumber\\
 &G_4=\left[h_3+(2\pi\ell_p)^3
 \left(N_5^+e_{3,+}+N_5^-e_{3,-}\right)\right]\wedge s_z,
\end{align}
where $h_3=\dd\mathcal B_2-(2\pi\ell_p)^3
(N_5^+\Theta_{3,+}+N_5^-\Theta_{3,-})$.
The descendant $\Theta_{3,\pm}$ is the difference of the
left and right Chern--Simons three-forms of these connections,
\begin{equation}\label{eq:theta-definition}
 \Theta_{3,\pm}=\frac{1}{8\pi^2}
 \left[\operatorname{CS}_3(\mathcal A_{\pm,R})
       -\operatorname{CS}_3(\mathcal A_{\pm,L})\right],
 \qquad \dd\operatorname{CS}_3(\mathcal A)=\tr(\mathcal F^2).
\end{equation}
Here $\operatorname{CS}_3(\mathcal A)=
\tr(\mathcal A\dd\mathcal A+\frac23\mathcal A^3)$,
in the anti-Hermitian fundamental trace convention.
Thus $\dd(e_{3,\pm}-\Theta_{3,\pm})=0$,
as required by the magnetic Bianchi identity.
The electric Maxwell flux through
$\Sigma_7=S^3_+\times S^3_-\times S^1_w$ is
$Q_2=(2\pi\ell_p)^6N_2$. Its conjugate momentum is
$\Pi_{\rm el}=Q_2/(2\kappa_{11}^2)$, where
$2\kappa_{11}^2=(2\pi)^8\ell_p^9$.
To specify the electric ensemble, let
$S_{\rm el}=-\frac12\int\mathcal Z h_3\wedge *_3h_3$.
The boundary variation of this kinetic term fixes the
two-form potential $\mathcal B_2$. Fixing the electric momentum instead
requires the Legendre term,
\begin{align}\label{eq:page-ensemble-variation}
 S_{\rm fix}&=S_{\rm el}
  -\int_{\partial\AdS_3}\Pi_{\rm el}\mathcal B_2,\nonumber\\
 \left.\delta S_{\rm el}\right|_{\partial}
 &=\int_{\partial\AdS_3}\Pi_{\rm el}\,\delta\mathcal B_2,
 &\left.\delta S_{\rm fix}\right|_{\partial}
 &=-\int_{\partial\AdS_3}\mathcal B_2\,\delta\Pi_{\rm el}.
\end{align}
Here the classical Page and Maxwell fluxes coincide on the
charge cycle. In first-order form, the charge-fixed action is
\begin{align}\label{eq:main-first-order-electric}
 S_{\rm fix}^{(1)}={}&-\frac12\int_{\AdS_3}
           \mathcal Z h_3\wedge *_3h_3
 -\int_{\AdS_3}\Pi_{\rm el}
   \left(h_3-\dd\mathcal B_2
   +(2\pi\ell_p)^3\sum_\pm N_5^\pm\Theta_{3,\pm}\right)
   \nonumber\\
 &-\int_{\partial\AdS_3}\Pi_{\rm el}\mathcal B_2.
\end{align}
Varying $\mathcal B_2$ gives $\dd\Pi_{\rm el}=0$.
The $\dd\mathcal B_2$ term cancels the Legendre boundary
term after integration by parts. Flux quantization and the angular-form
normalization give
\begin{equation}\label{eq:page-fixed-cs}
 S_{\rm CS}=-\Pi_{\rm el}(2\pi\ell_p)^3
 \sum_{\pm}N_5^\pm\int_{\AdS_3}\Theta_{3,\pm},\qquad
 \Pi_{\rm el}(2\pi\ell_p)^3N_5^\pm=2\pi N_2N_5^\pm.
\end{equation}

Substitution of the normalized angular-form descent
\eqref{eq:theta-definition} into the Page-fixed coupling
\eqref{eq:page-fixed-cs} gives the three-dimensional action

\begin{align}\label{eq:page-fixed-level-step}
 S_{\rm CS}
 &=-\sum_{\pm}\frac{\Pi_{\rm el}(2\pi\ell_p)^3N_5^\pm}{8\pi^2}
 \int_{\AdS_3}\bigl[\operatorname{CS}_3(\mathcal A_{\pm,R})
 -\operatorname{CS}_3(\mathcal A_{\pm,L})\bigr]\nonumber\\
 &=-\sum_{\pm}\frac{N_2N_5^\pm}{4\pi}
 \int_{\AdS_3}\bigl[\operatorname{CS}_3(\mathcal A_{\pm,R})
 -\operatorname{CS}_3(\mathcal A_{\pm,L})\bigr].
\end{align}

The coefficients determine the $SU(2)$ chiral levels
in \eqref{eq:bulk-levels}.
For a Hopf connection $\mathcal A=\mathrm i\sigma_3 a$,
$\tr(\mathcal A\dd\mathcal A)=-2a\dd a$;
the Abelian level is twice the corresponding $SU(2)$ level.
This derivation follows the angular-form method of
\cite{HansenKraus,HST2025}, keeping the electric
polarization explicit.

\section{Spinor data for the compact solutions}\label{app:spinors}

We use the eleven-dimensional Killing-spinor equation
\eqref{eq:11d-kse} in the conventions of
\cite{Ashmore,GNS}.

{For $S^3_+\times S^3_-\times T^2$ the radius relation
reads $(L/R_+)^2+(L/R_-)^2=1$, and the Hopf coframes obey
$\dd e^3=2(L/R_+)e^{12}$ and $\dd e^6=2(L/R_-)e^{45}$.}
A compatible internal spinor is obtained from the projector
\begin{equation}
 \Pi=\tfrac1{16}(1-\gamma_7)
 (1-i\gamma_{12})(1-i\gamma_{45})
 {[1+i((L/R_-)\gamma_3-(L/R_+)\gamma_6)\gamma_8]} .
\end{equation}
Its nonzero bilinears relevant to the volume are
\begin{align}
 K&=-e^7,&
 \xi^\flat&={\tfrac12\Bigl(\tfrac{L}{R_+}e^3+\tfrac{L}{R_-}e^6\Bigr)},\nonumber\\
 \omega&={-\Bigl(\tfrac{L}{R_+}e^3+\tfrac{L}{R_-}e^6\Bigr)\wedge e^7},&
 J&={e^{12}+e^{45}-\Bigl(\tfrac{L}{R_-}e^3-\tfrac{L}{R_+}e^6\Bigr)\wedge e^8},\nonumber\\
 \Psi&=-\tfrac12J\wedge J,&
 \eta&={2\Bigl(\tfrac{L}{R_+}e^3+\tfrac{L}{R_-}e^6\Bigr)},\nonumber\\
 \Omega&={4\Bigl(\tfrac{L^2}{R_+^2}e^{12}+\tfrac{L^2}{R_-^2}e^{45}\Bigr)},&
 \eta\wedge\Omega\wedge\Psi
 &=-8e^{1234568}.
 \label{eq:app-spheres-forms}
\end{align}
The final relation gives the normalization
in \eqref{eq:spheres-local}.
{For the explicit wedge computation, use the radius
relation,}
\begin{align}\label{eq:spheres-wedge-step}
 \eta\wedge\left(\frac{L}{R_-}e^3-\frac{L}{R_+}e^6\right)
 &=-2e^{36},\nonumber\\
 \Omega\wedge(e^{12}+e^{45})&=4e^{1245},\nonumber\\
 \Psi&=-e^{1245}
 +(e^{12}+e^{45})\wedge
 \left(\frac{L}{R_-}e^3-\frac{L}{R_+}e^6\right)\wedge e^8.
\end{align}
The first term in $\Psi$ wedges to zero with $\Omega$.
The remaining term gives
$\eta\wedge\Omega\wedge\Psi=-8e^{1234568}$,
with the sign inherited from the GNS internal orientation.

For clarity, the metric dual and the vector generator
are normalized separately. In period-$4\pi$ Hopf angles
$\psi_\pm$, choose
$H_\pm=2\partial_{\psi_\pm}$ and
$\eta_\pm=\tfrac12(\dd\psi_\pm+\cos\theta_\pm
\dd\phi_\pm)$. Then $\eta_\pm(H_\pm)=1$ and
$g_{M_8}(H_\pm,\cdot)=R_\pm^2\eta_\pm/L^2$.
The spinor one-form in \eqref{eq:app-spheres-forms}
is $\xi^\flat=(\eta_++\eta_-)/2$;
raising the index gives exactly
\eqref{eq:spinor-hopf-generator}. The chosen complex
supercharge has $H_+$ and $H_-$ charges $+1$, and thus
$R=2\xi$ has unit supercharge charge.

\subsection{The K3 factor}\label{app:k3-curvature}

Let $j$ be the parallel K\"ahler form, with
$\vol_{K3}=j^2/2=e^{4568}$ and
$\nabla^{K3}\chi=0$. The product spinor has
\begin{align}
 K&=-e^7,&\xi^\flat&=\tfrac12 e^3,&
 \omega&=-e^{37},\nonumber\\
 J&=e^{12}+j,&
 \Psi&=-e^{12}\wedge j-\vol_{K3},&
 \eta&=2e^3,\qquad\Omega=4e^{12}.
\end{align}

The terms containing $e^{12}\wedge j$ vanish after
wedging with $\Omega$, while the K3 volume form supplies
the remaining four horizontal directions. Thus

\begin{equation}\label{eq:k3-wedge-step}
 \eta\wedge\Omega\wedge\Psi
 =-8e^{123}\wedge\vol_{K3}.
\end{equation}

With $\vol_8=e^{12345678}$,
\begin{equation}
 e^{1237}\wedge\vol_{K3}=-\vol_8,\quad
 \int_{K3}p_1=3\sigma(K3)=-48,\quad
 \int_{K3}\frac14p_1=-12.
\end{equation}

The closed two-form freedom cannot change the electric
period in this example. Indeed, the cohomology of the
product satisfies
$H^2(S^3\times K3\times S^1_z;\mathbb R)
\cong H^2(K3;\mathbb R)$. Any invariant closed
$\vartheta_2$ therefore differs from a pullback of a
closed K3 form by $\dd\Lambda$. Averaging $\Lambda$
over the compact Hopf circle makes it invariant. Since
$\xi$ acts only on $S^3$, Cartan's identity then gives
$\iota_\xi\vartheta_2=-\dd(\iota_\xi\Lambda)$ and
$\oint_{S^1_z}\iota_\xi\vartheta_2=0$. Thus the
magnetically reconstructed period is independent of the
representative allowed in
\eqref{eq:magnetic-representative-shift}.

One can also verify the anomaly directly on the
seven-dimensional quotient
$B_7=S^2\times K3\times S^1_z$. Here the $\xi$ flow
period is $\ell_\xi=4\pi$, the normalized circle-bundle
Euler class obeys $\int_{S^2}e=1$, and the electric
one-form has $\int_{S^1_z}\mathrm e^{3\lambda}f
=2L^2\mathcal L_z$. The quotient orientation fixed by
the eight-dimensional frame gives

\begin{equation}\label{eq:k3-electric-anomaly-pairing}
 \int_{B_7}e\wedge\overline{\mathrm e^{3\lambda}f}
 \wedge p_1(TB_7)
 =-(2L^2\mathcal L_z)(-48)
 =96L^2\mathcal L_z.
\end{equation}

Substitution into the free-circle formula
\eqref{eq:anomaly-electric-class} yields

\begin{align}\label{eq:k3-anomaly-reduction}
 \left.(c_L-c_R)\right|_{C_3X_8}
 &=-\frac{192\pi^2L^2\mathcal L_z}
 {4(2\pi\ell_p)^3}\nonumber\\
 &=-\frac{6L^2\mathcal L_z}{\pi\ell_p^3}
 =-12N_5,
\end{align}

in agreement with the magnetic computation
\eqref{eq:k3-anomaly-calculation}.
These conventions fix the sign in
\eqref{eq:k3-anomaly-result}. The shifted
M-theory flux quantization
$[G_4/(2\pi\ell_p)^3]-p_1/4\in H^4(M;\mathbb Z)$
must be distinguished from the classical
electric flux used in \eqref{eq:k3-charge-integrals}.

\section{The periodic mixed BPS solution}\label{app:mixed-proof}

\subsection{Regularity at the caps}

For one summand of the periodic BPS function centered
at a real number $c$, direct differentiation gives
\begin{align}\label{eq:mixed-G-BPS-check}
 \partial_w\frac{w-c}{|w-c|}&=\frac1{2|w-c|},
 &\partial_w\log h&=\frac1{w-\bar w},\nonumber\\
 \frac{\mathrm i}{2}\frac{w-c}{|w-c|}
 +\text{c.c.}&=-\frac{v}{|w-c|},
 &\partial_w\left(\frac{\mathrm i}{2}
 \frac{w-c}{|w-c|}\right)
 &=\frac{\mathrm i}{4|w-c|}.
\end{align}
The locally convergent paired series therefore obeys
$\partial_wG=\frac12(G+\bar G)\partial_w\log h$.

For completeness, let $\chi(x)$ be the periodic indicator
of the black intervals of length $\nu T$. Its Fourier
coefficients, used in the exact bulk integral, are

\begin{equation}\label{eq:mixed-indicator}
 \chi_n=\frac1T\int_0^T
 \chi(x)\mathrm e^{-\ii k_nx}\,\dd x
 =\frac{\mathrm e^{-\ii\pi n\nu}\sin(\pi n\nu)}{\pi n},
 \quad k_n=\frac{2\pi n}{T}\ (n\ne0),\qquad
 \chi_0=\nu.
\end{equation}

Writing $s=|G|^2$ and $W_\pm=2(q\pm s)$,
the warp factors appearing in
\eqref{eq:periodicmetric} are

\begin{align}\label{eq:mixed-warp-factors}
 f_1^6&=\frac{4h^2(1-s)W_+}{W_-^2},&
 f_2^6&=\frac{4h^2(1-s)W_-}{W_+^2},
 \nonumber\\
 f_3^6&=\frac{h^2W_+W_-}{16(1-s)^2},&
 \rho^6&=\frac{L^{18}(1-s)W_+W_-}{16h^4}.
\end{align}

The positivity proof below ensures $W_\pm>0$
and smooth warp factors for $v>0$. Since
$\chi_1\ne0$, the solution is periodic in $x$
but does not possess a continuous $x$-translation
isometry.

Write $G=m+iq$ and $s=m^2+q^2$.
For the periodic interval indicator $\chi$,
\begin{equation}\label{eq:mixed-moment-bound}
 (m,q)=\frac12\int_{\mathbb R}\chi(t)
 \frac{((x-t)v,v^2)}{((x-t)^2+v^2)^{3/2}}\,\dd t,
 \qquad 0<s<q<1\quad(v>0).
\end{equation}
At fixed $q$, the ratio of the $m$ kernel to the
positive $q$ kernel in \eqref{eq:mixed-moment-bound} is
$(x-t)/v$, which is strictly monotone in $t$.
Rearrangement at fixed $q$ therefore makes an upper
or lower half-line extremal for $|m|$. For the half-line
$t>x+t_0v$, the extremal moments are
\begin{equation}\label{eq:mixed-moment-extremizer}
 {q=\frac12\left(1-\frac{t_0}{\sqrt{1+t_0^2}}\right),
 \qquad |m|=\frac{1}{2\sqrt{1+t_0^2}},}
 \qquad m^2=q(1-q).
\end{equation}
Since a periodic union of finite intervals is never a
half-line, it has $m^2<q(1-q)$ and thus $s<q<1$.
At the cap {$F_a$} put $w=re^{i\theta}$,
\begin{equation}\label{eq:mixed-cap-constant}
 \mathfrak c_{\rm cap}
 =-\frac{\psi(\nu)+\psi(1-\nu)+2\gamma_{\rm E}}{T}>0.
\end{equation}

\begin{equation}\label{eq:mixed-cap-G-expansion}
 G=\frac{\ii}{2}(1+\mathrm e^{\ii\theta})
 +\frac{\mathfrak c_{\rm cap}}2r\sin\theta+O(r^2).
\end{equation}

With $R^2=2^{4/3}L^2\mathfrak c_{\rm cap}^{1/3}r$,
\begin{align}
 \dd s_8^2&=\dd R^2+R^2\bigl[\dd(\theta/2)^2
 +\sin^2(\theta/2)\dd s_{S^3_2}^2
 +\cos^2(\theta/2)\dd s_{S^3_3}^2\bigr]+O(R^4)\nonumber\\
 &=\dd R^2+R^2\dd s_{S^7}^2+O(R^4).
\end{align}

In particular, $r_2=R\sin(\theta/2)$ and
$r_3=R\cos(\theta/2)$ are the radial coordinates
of the two $\mathbb R^4$ factors in
\eqref{eq:mixed-flat-cap}. The apparent boundary
of the half-cylinder is thus a smooth interior
point of the full sphere fibration at each cap.

For flux regularity, set $U=r+x$, $V=r-x$.
Here $\mathcal P$ is the auxiliary potential fixed by
$\partial_w\mathcal P=-\ii L^3\bar G$; the $b_2,b_3$ of
\eqref{eq:mixed-magnetic-potentials} are the magnetic
potentials, not independent fields.
Near $U=V=0$ the regular remainder admits
analytic functions {$m_1,q_1,\mathcal P_1$} such that
{%
\begin{align}
 m&=-\frac{\sqrt{UV}}{U+V}+\sqrt{UV}\,m_1(U,V),\nonumber\\
 q&=\frac{U}{U+V}+UV\,q_1(U,V),\nonumber\\
 \mathcal P&=\mathcal P_a-L^3U+UV\,\mathcal P_1(U,V),
 \qquad \mathcal P_1|_{UV=0}=L^3m_1|_{UV=0}.
\end{align}}%
Here $\mathcal P_a=\mathcal P(F_a)$.
The denominators
\begin{align}
 Z_3&=(1-s)\frac{U+V}{V},&
 Z_2&=(s+q)\frac{U+V}{U},\nonumber\\
 Z_3(0,V)&=1,&Z_2(U,0)&=2,\nonumber\\
 Z_1&={2m_1+(V-U)q_1-(U+V)(m_1^2+UVq_1^2)},
 &Z_1(0,0)&=\mathfrak c_{\rm cap}>0
\end{align}
are analytic and nonzero near the cap. Substitution
in the magnetic potentials of
\cite{GentleGutperleMarasinou} gives
\begin{align}
 b_3-b_{3,a}
 &={U\left[L^3\left(1-
   \frac{1-(U+V)m_1}{Z_3}\right)-V\mathcal P_1\right]},\nonumber\\
 b_2-b_{2,a}
 &={V\left[4L^3\frac{1-(U+V)m_1}{Z_2}
 -2L^3+2U\mathcal P_1\right]},\nonumber\\
 (b_1)_{\rm rat}
 &={4L^3\frac{1-(U+V)m_1}{Z_1}}.
\end{align}
Here $b_{i,a}=b_i(F_a)$, while
$(b_1)_{\rm rat}=2mh/(s-q)$ denotes the rational term
in the electric potential \eqref{eq:mixed-b1}.
{The first bracket vanishes at $U=0$ and the second at
$V=0$, by $Z_3(0,V)=1$, $Z_2(U,0)=2$ and the boundary condition on
$\mathcal P_1$. Hence $b_3-b_{3,a}$ is $U^2$ times an analytic
function and $b_2-b_{2,a}$ is $V^2$ times an analytic function.}
Since the respective
sphere radii squared are proportional to
$U,V$, the magnetic three-forms and
the electric one-form extend smoothly.

\subsection{\texorpdfstring{{Global spinors, signs and orientations}}{Global spinors, signs and orientations}}
\label{app:mixed-orientation}
\label{sec:periodic-spinor-audit}

The inverse Killing-spinor reconstruction of
\cite{DHEGKLocal} uses
\begin{equation}
 \varphi=\frac12\left[
 \operatorname{artanh}\left(
 -\frac{\sqrt{s}\,m}{q\sqrt{1-s}}\right)
 +i\arcsin\sqrt{s}\right],\qquad
 0<\operatorname{Im}\varphi<\frac\pi4.
\end{equation}
In this strip $\sinh\varphi$ and
$\cosh\varphi$ have continuous logarithms.
The reduced spinors can be chosen as
\begin{align}
 \alpha&=\left(\frac{L^3}{\rho}\right)^{1/4}
 |\cosh\varphi|^{1/2}e^{-i\arg(\cosh\varphi)},
 \nonumber\\
 \beta&=\left(\frac{L^3}{\rho}\right)^{1/4}
 |\sinh\varphi|^{1/2}e^{-i\arg(\sinh\varphi)},
 \qquad
 \zeta=-\frac{2\operatorname{Re}(\bar\alpha\beta)}
 {|\alpha|^2+|\beta|^2}.
\end{align}
{At the cap $F_a$ one has $\operatorname{Re}\varphi\to+\infty$
and $\alpha/\beta\to+1$, so $\zeta=-1$, and at $F_b$ one has
$\operatorname{Re}\varphi\to-\infty$ and $\alpha/\beta\to-1$, so
$\zeta=+1$.}
For completeness, the differential forms entering the direct
boundary integrals can be constructed from these two functions
in a fixed Clifford frame. In this representation the eight
internal Euclidean frame directions are relabeled $0,\ldots,7$,
rather than $1,\ldots,8$ as in Section~2; $\gamma_0$ is not a
Lorentzian time matrix. Let $\sigma_i$ be Pauli matrices
and choose the eight gamma matrices
{%
\begin{align}\label{eq:mixed-gamma-frame}
 (\gamma_0,\gamma_1,\gamma_2)&=(\sigma_2,\sigma_1,\sigma_3)
 \otimes\mathbf1\otimes\sigma_2\otimes\sigma_3,\nonumber\\
 (\gamma_3,\gamma_4,\gamma_5)&=\mathbf1\otimes(\sigma_2,\sigma_1,\sigma_3)
 \otimes\sigma_3\otimes\sigma_3,\nonumber\\
 \gamma_6&=\mathbf1\otimes\mathbf1\otimes\mathbf1\otimes\sigma_1,
 \qquad\gamma_7=\mathbf1\otimes\mathbf1\otimes\mathbf1\otimes\sigma_2.
\end{align}}%
With $\gamma_9=\gamma_0\cdots\gamma_7$, the unit spinor is
\begin{equation}\label{eq:mixed-explicit-spinor}
 \chi=\frac{(0_4,\,\bar\alpha,\alpha,-\bar\beta,\beta,\,0_8)^T}
 {\sqrt{2(|\alpha|^2+|\beta|^2)}}.
\end{equation}
The Hopf generator follows by actual matrix multiplication.
Restricted to the four nonzero slots of $\chi$, the
matrices entering the bilinear are
\begin{equation}\label{eq:mixed-gamma-generator-subblock}
 \left.-\frac{\mathrm i}{2}\gamma_9\gamma_2
 \right|_{4\text{ slots}}
 =\frac12\begin{pmatrix}
 1&0&0&0\\0&1&0&0\\0&0&-1&0\\0&0&0&-1
 \end{pmatrix},\qquad
 \left.-\frac{\mathrm i}{2}\gamma_9\gamma_5
 \right|_{4\text{ slots}}
 =\frac12\begin{pmatrix}
 0&0&-\mathrm i&0\\0&0&0&-\mathrm i\\
 \mathrm i&0&0&0\\0&\mathrm i&0&0
 \end{pmatrix}.
\end{equation}

Inserting $(\bar\alpha,\alpha,-\bar\beta,\beta)$
gives the two nonzero components

\begin{equation}\label{eq:mixed-spinor-components}
 -\frac{\ii}{2}\bar\chi\gamma_9\gamma_2\chi
 =-\frac{f_2}{2f_1},\qquad
 -\frac{\ii}{2}\bar\chi\gamma_9\gamma_5\chi
 =-\frac{f_3}{f_1}.
\end{equation}

The other six vector components vanish. Raising the index
with the internal metric then gives the coordinate-basis
formula \eqref{eq:mixed-spinor-generator-derivation}.
The signed orthonormal coframe used in these expressions is
$e^{0,1,2}=f_2\hat e^{0,1,2}$,
$e^{3,4,5}=f_3\hat e^{3,4,5}$,
$e^6=2\rho\,\dd x$ and $e^7=2\rho\,\dd v$,
with $\dd\hat e^2=-2\hat e^{01}$ and
$\dd\hat e^5=-2\hat e^{34}$. The signed radii satisfy
\begin{equation}
 f_1=|\alpha|^2+|\beta|^2,\qquad
 f_2=-(|\alpha|^2-|\beta|^2),\qquad
 f_3=-\operatorname{Im}(\bar\alpha\beta).
\end{equation}
The bilinear definitions \eqref{eq:bilinear-definitions}
together with \eqref{eq:mixed-magnetic-potentials}
and \eqref{eq:mixed-gamma-frame} now specify
$J,\Psi,\omega,{\mathrm e^{3\lambda}F}$ and hence the full seven-form
$Q_7(y)$ algebraically. The unit sphere volumes are
$2\pi^2$ in absolute value. The physical volume
orientation reverses the signed spinor-frame orientation,
as used in \eqref{eq:mixed-direct-boundary-integrals}.
The metric, flux and spinor bilinears are $T$-periodic;
with the periodic spin structure the reconstructed spinor
is globally defined. The local electric potentials have the
monodromy stated in the main text. The canonical
$\mathcal N=(2,0)$ R-symmetry generator is
\begin{equation}
 \xi=-\tfrac12 H_2-H_3,\qquad
 \operatorname{period}(\xi)=4\pi,\qquad
 {\zeta(F_a)=-1,\quad\zeta(F_b)=+1.}
\end{equation}
The two Hopf generators vanish together
at the smooth caps, whose normal weights
have magnitudes $(1/2,1/2,1,1)$.

The source solution in
\cite{GentleGutperleMarasinou} uses the
opposite three-sphere orientation to the
GNS Clifford frame. With
{$\widehat\vol_i=-\vol_i$, where $\vol_i$ carries the
orientation of \cite{GentleGutperleMarasinou},}
{%
\begin{equation}
 C_3=b_1\vol_{\AdS_3}
 -b_2\widehat\vol_2-b_3\widehat\vol_3,\qquad
 \mathrm e^{3\lambda}f=-\dd b_1,\qquad
 \mathrm e^{3\lambda}F=-\dd(b_2\widehat\vol_2+b_3\widehat\vol_3).
\end{equation}}%
The signed spinor frame has the opposite
orientation to the positive physical
warped volume. All fixed, monodromy and outer
contributions in Section~\ref{sec:mixed-periodic-main}
use the latter orientation.

\paragraph{{Fixed-point orientations and the cubic contact term.}}
\label{app:mixed-sign-details}

The signed normal weights at either cap are
{$(-\tfrac12,-\tfrac12,-1,-1)$, the Hopf coefficients of
$\xi=-\tfrac12H_2-H_3$, and their product is $1/4$.} The two cap
orientations differ because their shrinking
three-sphere radii are exchanged. With
$x-x_F=r\cos\theta$ and $v=r\sin\theta$,
\begin{align}\label{eq:mixed-cap-orientation-jacobians}
 F_a:&\quad(r_2,r_3)=
 \left(R\sin\frac\theta2,R\cos\frac\theta2\right),
 &\frac{\partial(r_2,r_3)}{\partial(R,\theta)}
 &=-\frac R2,\nonumber\\
 F_b:&\quad(r_2,r_3)=
 \left(R\cos\frac\theta2,R\sin\frac\theta2\right),
 &\frac{\partial(r_2,r_3)}{\partial(R,\theta)}
 &=+\frac R2.
\end{align}
{The complex orientation of $\mathbb C^2\times\mathbb C^2$
is $-r_2^3r_3^3\,\dd r_2\wedge\dd r_3\wedge\vol_{S^3_2}\wedge\vol_{S^3_3}$.
Comparing it with $\dd x\wedge\dd v\wedge\vol_{S^3_2}\wedge\vol_{S^3_3}$
shows that the complex orientation agrees with the coordinate
orientation at $F_a$ and is opposite at $F_b$. The
coordinate-oriented fixed-point residues are therefore}
{%
\begin{equation}
 I_{F_a}^{\mathrm{coord}}
 =\frac{(2\pi)^4}{192\cdot\frac14}\,y_{F_a}^3
 =\frac{\pi^4}{3}\,y_{F_a}^3,\qquad
 I_{F_b}^{\mathrm{coord}}=-\frac{\pi^4}{3}\,y_{F_b}^3 .
\end{equation}}%
The physical orientation reverses this signed frame
and yields the two-cap result
\eqref{eq:mixed-fixed-physical}.

For the cubic cut term, write
{$\eta=-2\tau\eta_2-(1-\tau)\eta_3$},
where the Hopf contact normalization is
$\int_{S^3_i}\eta_i\wedge\dd\eta_i=-4\pi^2$.
Its curvature and cubic top form are
{%
\begin{align}\label{eq:mixed-cubic-wedge-derivation}
 \dd\eta={}&-2\,\dd\tau\wedge\eta_2
 +\dd\tau\wedge\eta_3
 -2\tau\,\dd\eta_2
 -(1-\tau)\,\dd\eta_3,\nonumber\\
 \eta\wedge(\dd\eta)^3={}&
 24\,\tau(1-\tau)\,
 \dd\tau\wedge(\eta_2\wedge\dd\eta_2)
 \wedge(\eta_3\wedge\dd\eta_3).
\end{align}}%
The cube contains one $\dd\tau$ and one curvature
from each Hopf base since $(\dd\eta_i)^2=0$.

The contact normalizations give, in the coordinate
orientation of the cut $\mathcal C_Y$,

\begin{align}\label{eq:mixed-cubic-contact-integration}
 \int_{\mathcal C_Y}\eta\wedge(\dd\eta)^3
 &=384\pi^4\int_{\tau(x_{\rm cut},0)}^{\tau(x_{\rm cut},Y)}
      \tau(1-\tau)\,\dd\tau\nonumber\\
 &=64\pi^4
 \bigl[3\tau^2-2\tau^3\bigr]^{
   \tau(x_{\rm cut},Y)}_{\tau(x_{\rm cut},0)}.
\end{align}

Reversing to the positive physical warped-volume
orientation, the cubic part of the cut jump is

\begin{equation}\label{eq:mixed-cubic-finite}
 I_{\delta^3}(Y)=\frac{\pi^4\delta^3}{3}
 \left[
 3\tau(x_{\rm cut},Y)^2-2\tau(x_{\rm cut},Y)^3
 -3\tau(x_{\rm cut},0)^2+2\tau(x_{\rm cut},0)^3
 \right].
\end{equation}

On a white cut one has $\tau(x_{\rm cut},0)=1$;
on a black cut $\tau(x_{\rm cut},0)=0$.
The white cut has a free R-symmetry circle, whereas
the black cut has a $\mathbb Z_2$ stabilizer at its
lower edge because $\xi=-H_3$ there rather than the
generic $4\pi$-periodic generator.
Both choices define the same smooth eight-dimensional
integral when their outer-boundary pieces are
included. For the white cut the limit
$\tau(\infty)=(1-\nu)/(1+\nu)$ in
\eqref{eq:mixed-cubic-finite} produces the second
line of \eqref{eq:mixed-summary}.

\subsection{Boundary reduction in coordinate components}
\label{app:mixed-boundary-reproduction}

The regulated identity in
\eqref{eq:mixed-regulated-target} can be checked without
introducing further geometric fields. This appendix gives the
potential representatives and scalar coefficients needed to
evaluate the cut and outer integrals
\eqref{eq:mixed-direct-boundary-integrals} directly from the
periodic BPS solution. The intermediate coefficients below are
components of the already defined seven-form $Q_7(y)$.
Their direct check is the compact differential identity
\eqref{eq:mixed-explicit-curl}, followed by the exact
endpoint evaluation in Appendix~\ref{app:mixed-analytic-endpoints}.

In the sphere orientation of
Appendix~\ref{app:mixed-orientation}, the two magnetic
potentials are
\begin{equation}\label{eq:mixed-magnetic-potentials}
 b_2=-\frac{2mh}{s+q}+4L^3x+2\mathcal P,\qquad
 b_3=\frac{mh}{2(1-s)}-\mathcal P.
\end{equation}

The electric potential, in the same conventions, is

\begin{equation}\label{eq:mixed-b1}
 b_1=\frac{2mh}{s-q}+2\widetilde h-2\mathcal P,
 \qquad \widetilde h=2L^3x,\qquad
 \partial_w\mathcal P=-\ii L^3\bar G.
\end{equation}

Here $\widetilde h$ is the harmonic dual of $h$,
and $\mathcal P$ is an auxiliary potential rather than
an independent flux parameter. The spinorial scalar is

\begin{equation}\label{eq:mixed-scalar-potential}
 y=\frac12(f_1^3\zeta+b_1)+\sigma,
 \qquad
 \sigma\in\mathbb R.
\end{equation}

The bilinear equations give
$f_1^3\zeta=-2mh/(s-q)$; hence the rational
terms cancel \emph{exactly} and
$y=2L^3x-\mathcal P+\sigma$.
Differentiating the last expression reproduces
\eqref{eq:mixed-scalar-simplified}.
We normalize $\mathcal P(0,0)=0$ and retain $\sigma$
as the arbitrary additive electric branch constant.
The function $\mathcal P$ entering both the electric
potential \eqref{eq:mixed-scalar-potential} and these
magnetic potentials has the convergent Fourier
representation
\begin{equation}\label{eq:mixed-P-Fourier}
 \mathcal P=-2L^3\nu x-L^3T\nu(1-\nu)
 +4L^3\operatorname{Re}\sum_{n\geq1}
  \mathrm i\chi_n\mathrm e^{\mathrm ik_nx}
                   vK_1(k_nv),
\end{equation}
with $\partial_x\mathcal P=-2L^3q$ and
$\partial_v\mathcal P=2L^3m$.
The condition $\mathcal P(0,0)=0$, together with
$vK_1(k_nv)\to1/k_n$ and
$\sum_{n\ge1}\sin^2(\pi n\nu)/n^2
=\pi^2\nu(1-\nu)/2$, fixes the offset
$-L^3T\nu(1-\nu)$.

The magnetic potentials have the independent periods
$\Delta_xb_2=4L^3T(1-\nu)$ and
$\Delta_xb_3=2L^3T\nu$. On a sphere-cycle
$S_i^3\times S_x^1$, magnetic flux quantization is

\begin{equation}\label{eq:mixed-magnetic-quantization}
 |N_{5,i}|=\frac{1}{(2\pi\ell_p)^3}
 \left|\int_{S_i^3\times S_x^1}G_4\right|
 =\frac{2\pi^2|\Delta_xb_i|}{(2\pi\ell_p)^3}
 =\frac{|\Delta_xb_i|}{4\pi\ell_p^3}.
\end{equation}

For rational $\nu$ the overall scale
$L^3T/\ell_p^3$ may be chosen so that both
classical periods are integral.

For the boundary reduction, indices $0,1,2$ run
over the first unit three-sphere, $3,4,5$ over the
second, while $6,7$ denote the coordinate one-forms
$\dd x,\dd v$. A subscript $012345j$ extracts the
coefficient of
$\widehat{\vol}_2\wedge\widehat{\vol}_3
\wedge\dd x^j$, with
$\widehat{\vol}_i=-\vol_i$ in the spinor frame.
Their sphere integral is $4\pi^4$.
For the positive physical warped-volume orientation,
the cut and outer terms are

\begin{align}\label{eq:mixed-direct-boundary-integrals}
 I_{\rm monodromy}(Y)
 &=-4\pi^4\int_0^Y
   [\Delta_\delta Q_7]_{0123457}
                     (x_{\rm cut},v)\,\dd v,\nonumber\\
 I_{\rm outer}(Y)
 &=4\pi^4\int_{x_{\rm cut}}^{x_{\rm cut}+T}
   [Q_7]_{0123456}(x,Y)\,\dd x .
\end{align}

The second integral is taken over one periodic
window, with the cut as its lateral edge.
The orientation choice is the same as in the
fixed-point result \eqref{eq:mixed-fixed-physical}.

To separate the powers of the electric jump, expand the
same primitive as
{%
\begin{equation}\label{eq:mixed-Q-coefficients}
 Q_7(y)=Q_7^{(0)}+yQ_7^{(1)}+y^2Q_7^{(2)}+y^3Q_7^{(3)},
\end{equation}
where the four coefficients $Q_7^{(r)}$ are read
directly from \eqref{eq:Q7}.
The linear, quadratic and cubic cut terms are therefore
\begin{equation}\label{eq:mixed-jump-orders}
 \Delta_\delta Q_7=
 \delta\bigl(Q_7^{(1)}+2yQ_7^{(2)}+3y^2Q_7^{(3)}\bigr)
 +\delta^2\bigl(Q_7^{(2)}+3yQ_7^{(3)}\bigr)+\delta^3Q_7^{(3)}.
\end{equation}}%

We now reduce the boundary forms to scalar functions in
the coordinate frame specified immediately before
\eqref{eq:mixed-direct-boundary-integrals}. Here
$\partial_6=\partial_x$ and $\partial_7=\partial_v$.
The spinor bilinears reconstructed in
Appendix~\ref{app:mixed-orientation} are
\begin{align}\label{eq:mixed-boundary-spinor-scalars}
 J_{26}&=-\frac{2\operatorname{Re}(\alpha\beta)}{f_1},
 &J_{27}&=-\frac{2\operatorname{Im}(\alpha\beta)}{f_1},
 \nonumber\\
 J_{56}&=-\frac{\operatorname{Im}(\alpha^2+\beta^2)}{f_1},
 &J_{57}&=\frac{\operatorname{Re}(\alpha^2+\beta^2)}{f_1},
 \nonumber\\
 \omega_{26}&=J_{57},&\omega_{27}&=-J_{56},
 \nonumber\\
 \omega_{56}&=-J_{27},&\omega_{57}&=J_{26}.
\end{align}
Set $\mathsf C_{r,j}={[Q_7^{(r)}]_{012345j}}$, for $j=6,7$.
Wedge multiplication of \eqref{eq:Q7} gives
{%
\begin{align}\label{eq:mixed-scalar-C0}
 \mathsf C_{0,6}
 &=f_1^3\rho f_2^2f_3^2
 \Bigl[\tfrac12f_2(1-\tau)J_{57}
       +f_3\tau J_{27}\Bigr],\nonumber\\
 \mathsf C_{0,7}
 &=-f_1^3\rho f_2^2f_3^2
 \Bigl[\tfrac12f_2(1-\tau)J_{56}
       +f_3\tau J_{26}\Bigr].
\end{align}}%
{%
\begin{align}\label{eq:mixed-scalar-C1}
 \mathsf C_{1,j}
 =f_1^2\Bigl\{&
 \rho(1-\tau)
 \Bigl[\tfrac12f_2^3(1-\tau)
     -f_2f_3^2\tau\Bigr]J_{2j}
 \nonumber\\
 &+\rho\tau
 \Bigl[-f_2^2f_3(1-\tau)
     +2f_3^3\tau\Bigr]J_{5j}
 +\tfrac14f_2^2f_3^2
        \partial_j\tau\Bigr\},
\end{align}}%
{%
\begin{align}\label{eq:mixed-scalar-C2}
 \mathsf C_{2,j}
 ={}&\tfrac18(1-\tau)^2\partial_j b_2
 -\tfrac12\tau^2\partial_j b_3
 +\tfrac12f_1\zeta\Bigl[
 \tfrac12f_2^2(1-\tau)
 -f_3^2\tau\Bigr]\partial_j\tau
 \nonumber\\
 &-f_1\rho f_2
    \tau(1-\tau)^2\omega_{2j}
 +2f_1\rho f_3
    \tau^2(1-\tau)\omega_{5j},
\end{align}}%
{%
\begin{equation}\label{eq:mixed-scalar-C3}
 \mathsf C_{3,j}=-\frac{\tau(1-\tau)}{2}
   \partial_j\tau.
\end{equation}}%
Consequently the cut and outer integrands are
the scalar combinations
\begin{align}\label{eq:mixed-reproducible-boundary-functions}
 [Q_7]_{0123456}&=
 \mathsf C_{0,6}+y\mathsf C_{1,6}
 +y^2\mathsf C_{2,6}+y^3\mathsf C_{3,6},
 \nonumber\\
 [\Delta_\delta Q_7]_{0123457}&=
 \delta(\mathsf C_{1,7}+2y\mathsf C_{2,7}
                         +3y^2\mathsf C_{3,7})\nonumber\\
 &\quad+\delta^2(\mathsf C_{2,7}+3y\mathsf C_{3,7})
 +\delta^3\mathsf C_{3,7}.
\end{align}
These expressions involve only the BPS functions.
Substitution in
\eqref{eq:mixed-direct-boundary-integrals} determines
the three finite-cutoff monodromy integrals. Their exact
endpoint evaluation is given in
Appendix~\ref{app:mixed-analytic-endpoints}.

For the analytic $Y\to\infty$ limit, the magnetic
potential derivatives approach
$\partial_x b_2\to4L^3(1-\nu)$ and
$\partial_x b_3\to2L^3\nu$. Together with the limiting
spinor bilinears, they give
\begin{align}\label{eq:mixed-outer-integrand-limit}
 &\mathsf C_{0,6},\mathsf C_{1,6},\mathsf C_{3,6}
 \longrightarrow0,
 \qquad
 \mathsf C_{2,6}\longrightarrow
 L^3\frac{\nu(1-\nu)(3\nu-1)}{(1+\nu)^2},\nonumber\\
 &y(x,\infty)=y_{\rm cut}
  +\frac{\delta}{T}(x-x_{\rm cut}).
\end{align}
Inserting the resulting quadratic polynomial into
\eqref{eq:mixed-direct-boundary-integrals} and using
$\int_{x_{\rm cut}}^{x_{\rm cut}+T}y^2\dd x
=T[(y_{\rm cut}+\delta)^3-y_{\rm cut}^3]/(3\delta)$
gives the exact outer-boundary limit

\begin{equation}\label{eq:mixed-outer-limit}
 \begin{aligned}
 I_{\rm outer}(\infty)
 &=\frac{2\pi^4}{3}
 \frac{\nu(1-\nu)(3\nu-1)}{(1+\nu)^3}
 \left[(y_{\rm cut}+\delta)^3-y_{\rm cut}^3\right],
 \\
 y_{\rm cut}
 &=2L^3(1+\nu)x_{\rm cut}
   +L^3T\nu(1-\nu)+\sigma .
 \end{aligned}
\end{equation}

For $0<\nu<1$ this term is generically nonzero,
while it vanishes in the large-cutoff limit at
$\nu=1/3$. Its dependence on $x_{\rm cut}$ and
$\sigma$ cancels in the complete boundary sum.

\subsection{Analytic cut primitives and the finite-cutoff identity}
\label{app:mixed-analytic-endpoints}

\paragraph{Exact Fourier--Bessel bulk integral.}
The periodic image solution has the mode expansion
$q=\sum_nq_n(v)\mathrm e^{\ii k_nx}$ and
$m=\sum_nm_n(v)\mathrm e^{\ii k_nx}$, where
\begin{equation}\label{eq:mixed-Fourier-coefficients}
 \begin{aligned}
 q_n(v)&=\chi_n|k_n|vK_1(|k_n|v),\qquad
 m_n(v)=-\mathrm i\chi_nk_nvK_0(|k_n|v)\quad(n\ne0),\\
 q_0&=\nu,\qquad m_0=0.
 \end{aligned}
\end{equation}
Here $K_j$ are modified Bessel functions of the second kind;
$\chi_n$ and $k_n$ are given in \eqref{eq:mixed-indicator}.
Parseval's identity gives
\begin{align}\label{eq:mixed-parseval-step}
 \int_0^T(q-s)\,\dd x
 =T\biggl[\nu(1-\nu)
 -\sum_{n\ne0}|\chi_n|^2k_n^2v^2
 \Bigl(K_0(|k_n|v)^2+K_1(|k_n|v)^2\Bigr)\biggr].
\end{align}
The radial integral then follows from
\begin{equation}\label{eq:mixed-Bessel-primitive}
 \frac{\dd}{\dd z}\left[z^3K_0(z)K_1(z)+z^2K_1(z)^2\right]
 =-z^3\bigl[K_0(z)^2+K_1(z)^2\bigr],
\end{equation}
the limit $z^2K_1(z)^2\to1$ as $z\to0$, and the exact
sum $\sum_{n\ne0}|\chi_n|^2/k_n^2
=T^2\nu^2(1-\nu)^2/12$. These yield
the complete bulk expression

\begin{equation}\label{eq:mixed-analytic-bulk-volume}
 \begin{aligned}
 \mathcal V(Y)=16\pi^4L^9T\biggl[
 &\frac{\nu(1-\nu)Y^2}{2}
 -\frac{T^2\nu^2(1-\nu)^2}{12}\\
 &+\sum_{n\ne0}\frac{|\chi_n|^2}{k_n^2}
 \Bigl((|k_n|Y)^3K_0(|k_n|Y)K_1(|k_n|Y)\\
 &\hspace{8em}+(|k_n|Y)^2K_1(|k_n|Y)^2\Bigr)
 \biggr].
 \end{aligned}
\end{equation}

The series converges exponentially for every $Y>0$
and has the large-$Y$ expansion
\eqref{eq:mixed-bulk-large-Y}; no numerical integration
is required.

\paragraph{Rational boundary coefficients.}
The coefficients in \eqref{eq:mixed-scalar-C0}--\eqref{eq:mixed-scalar-C3}
admit an exact rational reduction. Substituting the spinors and
the potentials of the preceding subsection, and writing the
coordinate one-forms explicitly, gives
\begin{align}\label{eq:mixed-rational-boundary-coefficients}
 \mathsf C_{0,6}\dd x+\mathsf C_{0,7}\dd v
 &=-\frac{4L^9mv^2}{1+s+2q}
       \bigl[m\,\dd x+(1+q)\,\dd v\bigr],\nonumber\\
 \mathsf C_{1,6}\dd x+\mathsf C_{1,7}\dd v
 &=L^6\bigl\{v^2\dd\tau
   -v\tau(1-\tau)\bigl[m\,\dd x+(1+q)\,\dd v\bigr]\bigr\},
   \nonumber\\
 \mathsf C_{2,6}\dd x+\mathsf C_{2,7}\dd v
 &=L^3\dd x-\frac{1+3\tau^2-2\tau^3}{4}\dd y
   -L^3\dd\!\left(\frac{mv(1+3s+4q)}{2(1+s+2q)^2}\right),
   \nonumber\\
 \mathsf C_{3,6}\dd x+\mathsf C_{3,7}\dd v
 &=-\frac1{12}\dd(3\tau^2-2\tau^3).
\end{align}
These are the scalar components of the original
spinorial seven-form, rather than representatives inferred
from a known bulk volume. In particular, the identity
follows directly by differentiating
\eqref{eq:mixed-rational-boundary-coefficients},
using \eqref{eq:mixed-scalar-simplified} and the
BPS equations
$m_v=q_x+m/v$ and $q_v=-m_x$:

\begin{equation}\label{eq:mixed-explicit-curl}
 \partial_v[Q_7]_{0123456}
 -\partial_x[Q_7]_{0123457}
 =4L^9v(q-q^2-m^2).
\end{equation}

The terms proportional to $y$, $y^2$ and $y^3$
cancel separately. With the orientations in
\eqref{eq:mixed-direct-boundary-integrals}, this
establishes \eqref{eq:mixed-boundary-derivative}
directly from the explicit BPS data.

At a fixed cut away from the caps, the Fourier--Bessel
solution has $q\to\nu$, $m\to0$, and
$\tau\to(1-\nu)/(1+\nu)$ exponentially. The exact
coefficient forms therefore imply
\begin{align}\label{eq:mixed-asymptotic-coefficients}
 \mathsf C_{1,7}
 &=-\frac{2L^6\nu(1-\nu)}{1+\nu}\,v
  +O\!\left(v^{5/2}e^{-2\pi v/T}\right),\nonumber\\
 \mathsf C_{2,7}
 &=O\!\left(v^{3/2}e^{-2\pi v/T}\right),\nonumber\\
 \mathsf C_{3,7}
 &=O\!\left(v^{1/2}e^{-2\pi v/T}\right).
\end{align}
In particular, $\mathsf C_{2,7}$ has exactly vanishing
zero mode. The scalar $y$ has a finite limit at fixed
$x_{\rm cut}$, so only the term linear in $\delta$ carries
the quadratic divergence. Integrating
the separate cut orders gives

\begin{align}\label{eq:mixed-asymptotic-cut-orders}
 I_\delta(Y)
 &=8\pi^4L^9T\nu(1-\nu)Y^2+C_\delta
  +O\!\left(Y^{5/2}\mathrm e^{-2\pi Y/T}\right),\nonumber\\
 I_{\delta^2}(Y)
 &=C_{\delta^2}
  +O\!\left(Y^{3/2}\mathrm e^{-2\pi Y/T}\right),\nonumber\\
 I_{\delta^3}(Y)
 &=C_{\delta^3}
  +O\!\left(Y^{1/2}\mathrm e^{-2\pi Y/T}\right).
\end{align}

The finite constants are evaluated by the exact
endpoint primitives below. The absence of a
$\mathsf C_{2,7}$ zero mode is essential to the
finiteness of $I_{\delta^2}$.

The integrands of the three separate cut orders possess
explicit primitives. Define \(\mathcal H(x,v)\) by
\begin{equation}\label{eq:mixed-H-primitive}
 \dd\mathcal H
 =2L^3(y+L^3vm)\,\dd x-2L^6v(1-q)\,\dd v.
\end{equation}
Its integrability is an immediate consequence of the BPS
equations. For the periodic solution of
\eqref{eq:mixed-P-Fourier}, the normalized Fourier
representative is
\begin{align}\label{eq:mixed-H-Fourier}
 \mathcal H(x,v)={}&2L^6(1+\nu)x^2
 +2L^3\bigl[L^3T\nu(1-\nu)+\sigma\bigr]x
 -L^6(1-\nu)v^2\nonumber\\
 &-4L^6v^2\operatorname{Re}\sum_{n\geq1}
 \chi_n\mathrm e^{\mathrm ik_nx}K_2(k_nv).
\end{align}
With no further assumptions, the three primitives can be taken as
\begin{align}\label{eq:mixed-three-endpoint-primitives}
 \mathcal F_3={}&-\frac{3\tau^2-2\tau^3}{12},\nonumber\\
 \mathcal F_2={}&L^3x-\frac y4(1+3\tau^2-2\tau^3)
 -\frac{L^3mv(1+3s+4q)}{2(1+s+2q)^2},\nonumber\\
 \mathcal F_1={}&-\frac{y^2}4(1+3\tau^2-2\tau^3)
 -\frac{L^3ymv(1+3s+4q)}{(1+s+2q)^2}
 +L^6v^2\tau+\mathcal H(x,v).
\end{align}
Direct differentiation establishes the three one-form identities
\begin{align}\label{eq:mixed-endpoint-derivatives}
 \dd\mathcal F_3
 &=\mathsf C_{3,6}\dd x+\mathsf C_{3,7}\dd v,\nonumber\\
 \dd\mathcal F_2
 &=\sum_{j=6,7}(\mathsf C_{2,j}+3y\mathsf C_{3,j})\,\dd x^j,
 \nonumber\\
 \dd\mathcal F_1
 &=\sum_{j=6,7}(\mathsf C_{1,j}+2y\mathsf C_{2,j}
   +3y^2\mathsf C_{3,j})\,\dd x^j,
\end{align}
where \(\dd x^6=\dd x\) and \(\dd x^7=\dd v\).
Consequently the three physical cut integrals are exactly
\begin{equation}\label{eq:mixed-cut-endpoint-integrals}
 \boxed{\displaystyle
 I_{\delta^r}(Y)=-4\pi^4\delta^r
 \bigl[\mathcal F_r(x_{\rm cut},Y)
       -\mathcal F_r(x_{\rm cut},0)\bigr],\qquad r=1,2,3.}
\end{equation}
This evaluates each order separately for every finite cutoff
and does not use the bulk volume to determine an integration
constant.

The lower endpoint in \eqref{eq:mixed-cut-endpoint-integrals}
is also analytic. Let
\begin{equation}\label{eq:mixed-Bernoulli-cubic}
 B_3(t)=t^3-\frac32t^2+\frac12t,
 \qquad
 \Delta B_3=B_3(\{x_{\rm cut}/T\})
 -B_3(\{x_{\rm cut}/T-\nu\}),
\end{equation}
with \(B_3\) periodically continued, and set
\(\tau_\infty=(1-\nu)/(1+\nu)\). The Fourier limit
\(v^2K_2(k_nv)\to2/k_n^2\) gives
\begin{equation}\label{eq:mixed-H-lower-endpoint}
 \mathcal H(x,0)=2L^6(1+\nu)x^2
 +2L^3\bigl[L^3T\nu(1-\nu)+\sigma\bigr]x
 -\frac23L^6T^2
 \bigl[B_3(\{x/T\})-B_3(\{x/T-\nu\})\bigr].
\end{equation}
Write \(y_{\rm cut}=\lim_{v\to\infty}y(x_{\rm cut},v)\),
\(y_0=y(x_{\rm cut},0)\), and let \(S_0=1\) or \(0\)
according as the cut lies in a white or black interval.
Taking both endpoints of \eqref{eq:mixed-three-endpoint-primitives}
yields the complete finite terms of
\eqref{eq:mixed-asymptotic-cut-orders}:
\begin{align}\label{eq:mixed-cut-finite-constants}
 C_\delta={}&\pi^4\delta\left[
 \bigl(1+3\tau_\infty^2-2\tau_\infty^3\bigr)y_{\rm cut}^2
 -(1+S_0)y_0^2-\frac83L^6T^2\Delta B_3\right],\nonumber\\
 C_{\delta^2}={}&\pi^4\delta^2\left[
 \bigl(1+3\tau_\infty^2-2\tau_\infty^3\bigr)y_{\rm cut}
 -(1+S_0)y_0\right],\nonumber\\
 C_{\delta^3}={}&\frac{\pi^4\delta^3}{3}
 \bigl(3\tau_\infty^2-2\tau_\infty^3-S_0\bigr).
\end{align}
For either type of cut, substituting these constants, the
cap residues and the outer limit
\eqref{eq:mixed-outer-limit} cancels the cut location and
the branch parameter \(\sigma\) identically, giving

\begin{equation}\label{eq:mixed-asymptotic-constant-sum}
 \boxed{\displaystyle
 I_{\rm fixed}+C_\delta+C_{\delta^2}
 +C_{\delta^3}+I_{\rm outer}(\infty)
 =-\frac43\pi^4L^9T^3\nu^2(1-\nu)^2.}
\end{equation}

This independently reproduces the finite large-$Y$
coefficient of \eqref{eq:mixed-bulk-large-Y}
using boundary data alone.

It remains to fix the constant in the all-cutoff identity
without invoking the general Stokes theorem. From the last
line of \eqref{eq:mixed-rational-boundary-coefficients},
\begin{equation}\label{eq:mixed-cap-concentration}
 \mathsf C_{3,6}=-\frac1{12}\partial_x(3\tau^2-2\tau^3).
\end{equation}
As \(v\to0^+\), \(\tau\) tends to \(1\) on white
intervals and \(0\) on black intervals. On the chosen
periodic window the cubic coefficient therefore
concentrates with signed weights \(+1/12\) at \(F_a\)
and \(-1/12\) at \(F_b\). The other coefficients
have no cap point masses: near either cap, in terms of
\(r=\sqrt{(x-x_F)^2+v^2}\), they are respectively
\(\mathsf C_{0,6}=O(r^2)\),
\(\mathsf C_{1,6}=O(r)\), and
\(\mathsf C_{2,6}=O(1)\). Since \(y\) extends
continuously across the caps,
\begin{equation}\label{eq:mixed-cap-outer-cancellation}
 \lim_{Y\to0^+}I_{\rm outer}(Y)
 =\frac{\pi^4}{3}(y_{F_a}^3-y_{F_b}^3)
 =-I_{\rm fixed}.
\end{equation}
At a cut away from both caps, all three jump integrands
are \(O(v^3)\) and hence \(I_{\rm monodromy}(Y)=O(Y^4)\).
The boundary sum therefore tends to zero, as does
\(\mathcal V(Y)\). Combined with the independently
verified differential identity \eqref{eq:mixed-explicit-curl},
this proves \eqref{eq:mixed-all-cutoffs} for every \(Y>0\).

\end{document}